\RequirePackage{snapshot}
\documentclass{WileyMSP-template}
\usepackage{cmap}
\usepackage[utf8]{inputenc}
\usepackage[T1]{fontenc}
\usepackage{microtype}

\usepackage[unicode,pdfa]{hyperref}
\usepackage{hyperxmp}
\hypersetup{citecolor=black,
            colorlinks=false, 
            hidelinks,
            keeppdfinfo,
            linkcolor=black,
            pdfapart=3,
            pdfaconformance=B,
            pdfauthor={Umutcan Bektas, Maciej O. Liedke, Huan Liu, Fabian Ganss, Maik Butterling, Nico Klingner, Rene Hübner, Ilja Makkonen, Andreas Wagner,  Gregor Hlawacek},
            pdfcaptionwriter={Gregor Hlawacek},
            pdfcontactemail={g.hlawacek@hzdr.de},
            pdfkeywords={Gallium Oxide, Defects, Positron Annihilation Lifetime Spectroscopy},
            pdflang={en},
            pdfmetalang={en},
            pdfsubject={Polymorph conversion in Ga2O3},
            pdftitle={Defect analysis of the beta- to gamma-Ga2O phase transition},
            urlcolor=black
            }

\AtBeginDocument{%
  \hypersetup{%
    pdfpagerange={1-\ref*{TotPages}}
  }%
}
\immediate\pdfobj stream attr{/N 3^^J/Alternate/DeviceRGB} file{sRGB.icc}
\pdfcatalog{
  /PageMode /UseNone
  /OutputIntents [
    <<
      /Type /OutputIntent
      /S /GTS_PDFA1
      /DestOutputProfile \the\pdflastobj\space 0 R
      /OutputConditionIdentifier (sRGB IEC61966-2.1 (Equivalent to www.srgb.com 1998 HP profile))
      /Info(sRGB IEC61966-2.1 (Equivalent to www.srgb.com 1998 HP profile))
      /RegistryName (http://www.color.org/)
    >>
  ]
}
\usepackage[type={CC},
            modifier={by},
            version={4.0}]{doclicense}

\usepackage[numbers,sort&compress]{natbib}
\usepackage[labeled,resetlabels]{multibib}% If the numbering in the Suppl should restart with 1 add resetlabels to the options of multibib; remove the labeled option to remove the Suppl thing which is in the second ref list
\newcites{S}{Supplementary references}
\usepackage{xparse}
\let\oriCiteS\citeS

\RenewDocumentCommand{\citeS}{O{} O{} m}{%
  \renewcommand{\citenumfont}[1]{S##1}%  
  \oriCiteS[#1][#2]{#3}%
  \renewcommand{\citenumfont}[1]{##1}%
}
\usepackage{siunitx}
\DeclareSIUnit{\ions}{ion}
\DeclareSIUnit{\dpa}{dpa}

\usepackage{graphicx}
\usepackage{booktabs}
\usepackage{multirow}

\usepackage{miller}

\usepackage[nolist]{acronym}

\usepackage{chemformula}
\newcommand{\GaO}{\ch{Ga2O3}}
\newcommand{\BGaO}{$\beta$--\ch{Ga2O3}}
\newcommand{\GGaO}{$\gamma$--\ch{Ga2O3}}
\newcommand{\Ep}{$E_p$}

\usepackage[english]{babel}
\usepackage[normalem]{ulem}

\begin{document}

\begin{acronym}
  \acro{cbed}[CBED]{convergent beam electron diffraction}
  \acro{dbvepas}[DB-VEPAS]{Doppler broadening variable energy positron annihilation spectroscopy}
  \acro{dft}[DFT]{density functional theory}
  \acro{dpa}[dpa]{displacements per atom}
  \acro{ebsd}[EBSD]{electron backscatter diffraction}
  \acro{elbe}[ELBE]{electron linac for beams with high brilliance and low emittance}
  \acro{fft}[FFT]{fast Fourier transform}
  \acro{fib}[FIB]{focused ion beam}
  \acro{hzdr}[HZDR]{Helmholtz-Zentrum Dresden-Rossendorf}
  \acro{hrtem}[HRTEM]{high-resolution \acl*{tem}}
  \acro{meps}[MePS]{mono-energetic positron source}
  \acro{vepals}[VEPALS]{variable energy positron annihilation lifetime spectroscopy}
  \acro{pas}[PAS]{positron annihilation spectroscopy}
  \acro{rbs}[RBS]{Rutherford backscattering spectrometry}
  \acro{rbsc}[RBS/c]{\acl*{rbs} in channeling mode}
  \acro{tem}[TEM]{transmission electron microscopy}
  \acro{saed}[SAED]{selected area electron diffraction}
  \acro{stem}[STEM]{scanning \acl*{tem}}
  \acro{sem}[SEM]{scanning electron microscopy}
  \acro{sponsor}[SPONSOR]{slow positron system of Rossendorf}
  \acro{uwb}[UWB]{ultrawide-bandgap}
  \acro{vasp}[VASP]{Vienna ab-initio simulation package}
  \acro{vepfit}[VEPfit]{variable energy positron fitting}
  \acro{xrd}[XRD]{X-ray diffraction}
  \acro{bfstem}[BF-STEM]{bright-field \acl*{stem}}
  \acro{tde}[TDE]{threshold displacement energy}
  \acro{pad}[PAD]{planar atomic density}
  \acro{apb}[APB]{antiphase boundary}
  \acroplural{apb}[APBs]{antiphase boundaries}
  \acro{cdb}[cDB]{coincidence Doppler broadening}
  \acro{rsm}[RSM]{reciprocal space mapping}
\end{acronym}

% From the Wiley template to add the logo
\pagestyle{fancy}
\rhead{\includegraphics[width=2.5cm]{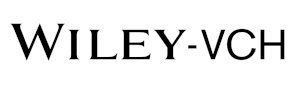}}

\title{Defect analysis of the $\beta$-- to \GGaO{} phase transition}
\maketitle

\author{Umutcan Bektas}
\author{Maciej O. Liedke}
\author{Huan Liu}
\author{Fabian Ganss}
\author{Maik Butterling}
\author{Nico Klingner}
\author{Ren\'e H\"ubner}
\author{Ilja Makkonen}
\author{Andreas Wagner}
\author{Gregor Hlawacek*}

\begin{affiliations}
  Umutcan Bektas, Fabian Ganss, Nico Klingner, Ren\'e H\"ubner, Gregor Hlawacek\\
  Institute of Ion Beam Physics and Materials Research, Helmholtz-Zentrum Dresden -- Rossendorf e.V., Bautzner Landstrasse 400, 01328 Dresden, Germany\\
  Email Address: u.bektas@hzdr.de; g.hlawacek@hzdr.de

  Maciej O. Liedke, Andreas Wagner\\
  Institute of Radiation Physics, Helmholtz-Zentrum Dresden -- Rossendorf e.V., Bautzner Landstrasse 400, 01328 Dresden, Germany
  
  Maik Butterling\\ Reactor Institute Delft, Delft University of Technology, Mekelweg 15, Nl-2629 JB Delft, The Netherlands

  Huan Liu, Ilja Makkonen\\
  Department of Physics, University of Helsinki, P.O.Box 43, FI-00014 University of Helsinki, Helsinki, Finland

\end{affiliations} 


\begin{abstract}
In this study, we investigate the ion irradiation induced phase transition in gallium oxide (\ch{Ga2O3}) from the $\beta$ to the $\gamma$ phase, the role of defects during the transformation, and the quality of the resulting crystal structure.
Using a multi-method analysis approach including \ac{xrd}, \ac{tem}, \ac{rbsc}, \ac{dbvepas} and \ac{vepals} supported by \ac{dft} calculations, we have characterized defects at all the relevant stages of the phase transition.
A reduction in backscattering yield was observed in \ac{rbsc} spectra after the transition to the $\gamma$ phase.
This goes hand in hand with a significant decrease in the positron trapping center density due to generation of embedded vacancies intrinsic for the \GGaO{} but too shallow in order to trap positrons.
A comparison of the observed positron lifetime of \GGaO{} with different theoretical models shows good agreement with the three-site $\gamma$ phase approach.
A characteristic increase in the effective positron diffusion length and the positron lifetime at the transition point from \BGaO{} to \GGaO{} enables visualization of the phase transition with positrons for the first time.
Moreover, a subsequent reduction of these quantities with increasing irradiation fluence was observed, which we attribute to further evolution of the \GGaO{} and changes in the gallium vacancy density as well as relative occupation in the crystal lattice.
\end{abstract}

\acresetall

\section{Introduction}

Throughout the history of science and engineering, there has always been a drive to find new and superior materials to advance technology. 
This is also true for power electronics, which needs new materials and related research activities to ensure future developments.
Gallium oxide (\ch{Ga2O3}) is a promising candidate for future \ac{uwb} semiconductor materials for power electronic devices. 
It has two main advantages over the traditional \ac{uwb} semiconductor materials, such as silicon carbide and gallium nitride.
\ch{Ga2O3} does not only have a superior breakdown voltage, which allows to increase the device performance~\cite{Qiao2022}, but cost-effective large wafer production by melt growth methods is also possible with Ga$_2$O$_3$~\cite{Heinselman2022}.
Furthermore, the existence of different polymorphs of Ga$_2$O$_3$ makes it a compelling material for crystal structure engineering.
This might be achieved via fabrication of different polymorphic layers on a single wafer as well as engineering of nanostructures with different polymorphs.
However, most research has focused on the $\beta$ phase, which is the most chemically and thermally stable polymorph of Ga$_2$O$_3$~\cite{Pearton2018}.

Ion implantation is a well-established method for enhancing the properties of semiconductors, and several attempts have been made to dope \BGaO{}~\cite{Seyidov2022, Azarov2021, Sardar2022}. 
Low-fluence rare-earth heavy ion irradiation has been employed to enhance the luminescence properties of \BGaO{}; however, ion-induced damage and the resulting structural disorder have negatively impacted luminescence performance~\cite{Lopez2013, Sarwar2024, Lorenz2014}.
Recently, \emph{Azarov et al.} demonstrated the high radiation tolerance of double-polymorph $\gamma$/$\beta$ Ga$_2$O$_3$ structures~\cite{Azarov2023}. 
Furthermore, Azarov et al. demonstrated that the fabrication of multilayer $\gamma$/$\beta$ polymorph structures is possible through controlled dynamic annealing~\cite{Azarov2025}.
This property makes Ga$_2$O$_3$ a good candidate for future electronic devices in space and nuclear applications, where radiation tolerance is key to ensuring consistent performance of electrical devices over decades~\cite{Polyakov2024}.
Attempts were made to increase the n-type conductivity of Ga$_2$O$_3$, and success was reported for silicon~\cite{Sardar2022}, tin~\cite{Akaiwa2016}, and hydrogen doping~\cite{Polyakov2023, Islam2020}.
However, the desired high n-type conductivity has not yet been achieved due to existing compensation centers in Ga$_2$O$_3$, i.e., gallium mono-vacancies~\cite{McCluskey2020,Korhonen2015}.
% I rearranged the sentences in the above and below paragraph. Now (according to me) the top dicusses applications while the bottom discusses the phasetransformation process as such.

Therefore, it is crucial to study and understand the formation of gallium-vacancy-related defects of the various \GaO{} polymorphs in their pristine state as well as during the phase transitions.
The latter is relevant to the design of future radiation-tolerant electronic devices in which the polymorphs of \GaO{} would be exploited to create new functionality.
It has been observed that high-energy and high-fluence ion irradiation, beyond a certain threshold, can induce a phase transition from the $\beta$ phase to the $\gamma$ phase of Ga$_2$O$_3$ instead of amorphization~\cite{Azarov2023}.
According to recent studies, the $\beta$-to-$\gamma$ phase transition may restrain the amorphization mechanism~\cite{Zhao2025}. 
This transition competes with amorphization and is driven by increased structural disorder and accumulated strain, which promote favorable atomic rearrangements~\cite{Azarov2022, Azarov2023, Huang2023a}.
A key driving factor for the transition from the monoclinic $\beta$ phase to the defective spinel $\gamma$ phase is the preservation of a stable and similar oxygen sublattice~\cite{Zhao2025, He2024a, Azarov2023}.
\emph{Huang} et al. demonstrated the atomic-scale mechanism of this transition and concluded that the relaxation of point defects at high concentrations, driven by lattice strain, plays a critical role~\cite{Huang2023a}.
\emph{Fern\'andez} et al. also showed that the transition occurs independently of the implanted ion species~\cite{GarciaFernandez2022}.

Here, we present a multi-method analysis approach (\ac{xrd}, \ac{tem}, \ac{rbsc}, \ac{dbvepas} and \ac{vepals}, both supported by \ac{dft} calculations to better understand the role of atomic defects for the \BGaO{} to \GGaO{} phase transition.
Defect formation and polymorph conversion are triggered by ion beam irradiation with noble gas ions to exclude chemical influences on the observed effects.
Our results offer unique insights into the defects in ion-beam-induced \GGaO{} and contribute to a deeper understanding of its radiation tolerance mechanisms.

\section{Results and Discussion}
%------------------------------------------------------------------------------%

\begin{figure}[btp]
    \includegraphics[width=1\linewidth]{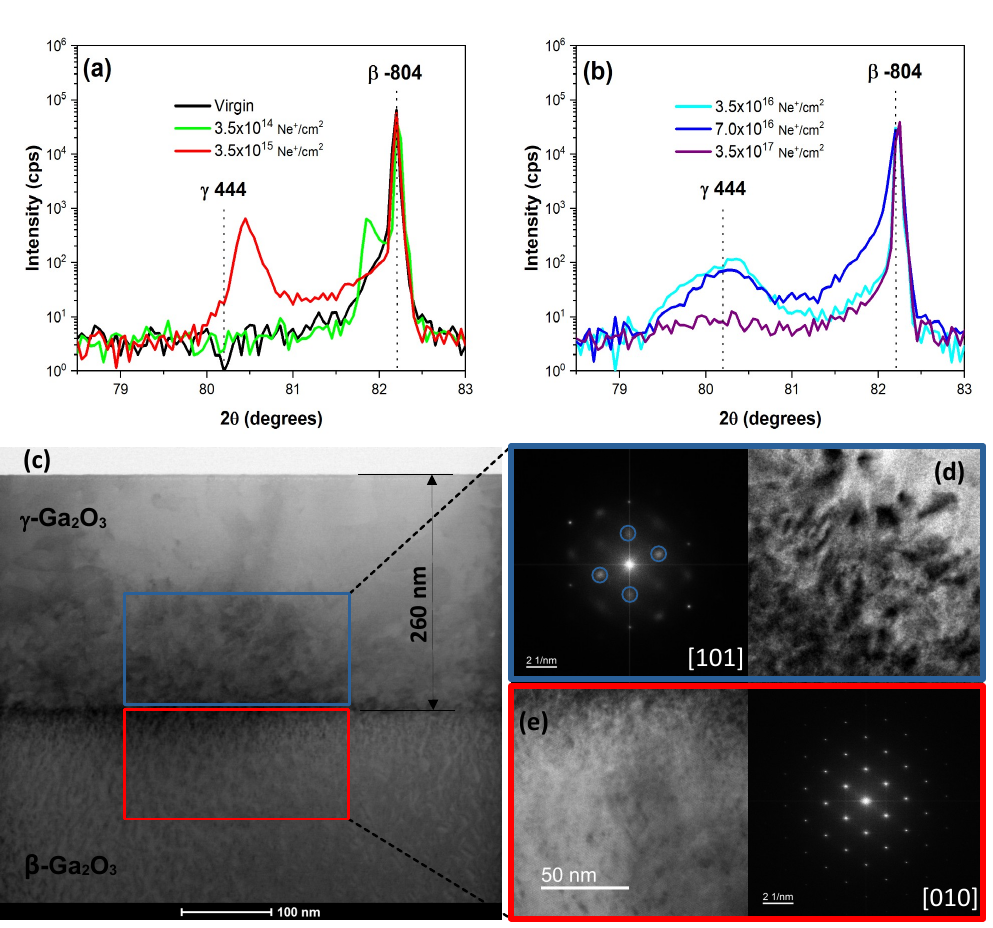}
    \caption{Structure analysis of \BGaO{} irradiated with different fluences of \qty{140}{keV} Ne$^+$.
    (a) \ac{xrd} patterns of the virgin and irradiated samples with low fluences,
    (b) \ac{xrd} patterns of the irradiated samples with high fluences.
    (c) \ac{bfstem} image of the converted layer on top of pristine \BGaO{}. 
    (d) \ac{hrtem} image (right) and \ac{fft} (left) of the $\gamma$ layer close to the interface along \hkl[101] zone axis. 
    The diffraction spots highlighted by the blue circles correspond to the \GGaO{}\,\hkl{111} reflections.
    (e) \ac{hrtem} image and \ac{fft} pattern of the $\beta$ region close to the interface along \hkl[010] zone axis.}
    \label{fig:Structure}
\end{figure}

In this study, commercial \hkl(-201)-oriented \BGaO{} substrates were irradiated with different fluences of \qty{140}{keV} Ne$^+$.
\ac{xrd} and \ac{tem} were performed to confirm the phase transition, as shown in Figure~\ref{fig:Structure}.
%-------------------------------------------------------------------
%-------------------------------------------------------------------

Based on the presented $\theta$/2$\theta$ scans in Figures~\ref{fig:Structure}(a) and (b) and the Supplementary Note~1, we conclude that the $\gamma$ phase formed with \hkl(111) orientation.
The 111, 222, 333, and 444 \GGaO{} reflections should usually be observable within a 2$\theta$ range from 10\textdegree{} to 90\textdegree{}.
However, only the $\gamma$\,222 and $\gamma$\,444 reflections are visible in our $\theta$/2$\theta$ scans. 
As discussed in Supplementary Note~1, this apparent inconsistency is attributed to the presence of \acp{apb} in the $\gamma$ phase~\cite{Yoo2022, GarciaFernandez2024, Tang2024}, which induce destructive interference, thereby suppressing the $\gamma$\,111 and $\gamma$\,333 reflections.
To focus on the $\gamma$\,444 reflection, only a small section of the \ac{xrd} patterns is presented here; the full diffractogram can be found in Supplementary Note~1. 

For all samples, the $\overline{8}04$ reflection of \BGaO{} at about 82.2\textdegree{} is expected due to diffraction from the unaffected substrate underneath the irradiated layer, whose thickness is less than the X-ray penetration depth.
If a $\gamma$ layer is present on top of the \BGaO{}, an additional peak near 80.2\textdegree{} is visible, which is identified as the 444 reflection of \GGaO{}.
As shown in Figure~\ref{fig:Structure}(a), after irradiation with a fluence of \qty{3.5e14}{\ions\per\centi\meter\squared}, an asymmetric shoulder of the $\overline{8}$04 reflection appears due to the lattice distortion and defect formation in the \BGaO{} crystal.
Increasing the fluence by one order of magnitude to \qty{3.5e15}{\ions\per\centi\meter\squared} results in a distinct peak at 80.5\textdegree{}, which is close to the characteristic 444 reflection of \GGaO{}.
The irradiation fluence corresponds to a peak damage of \qty{2.6}\,\ac{dpa} (see Supplementary Note~2), which is still one order of magnitude less than the \ac{dpa} required to induce the phase transition~\cite{Azarov2023}.
Therefore, we conclude that at a fluence of \qty{3.5e15}{\ions\per\centi\meter\squared}, the transition from the $\beta$ to the $\gamma$ phase has already started locally in small, highly damaged pockets but is not yet complete.

In Figure~\ref{fig:Structure}(b), after a further increase of the fluence by a factor of ten (\qty{3.5e16}{\ions\per\centi\meter\squared}), corresponding to a peak damage of \qty{26}{\dpa}, the 444 reflection of the $\gamma$ phase appears clearly at the expected position.
The broadening of the $\gamma$\,444 peak suggests the presence of inhomogeneous strain and lattice defects within the $\gamma$ layer.

It is important to note that the $\beta$-to-$\gamma$ phase transition occurs independently of the substrate's surface orientation~\cite{GarciaFernandez2022} and is, to a large extent, also independent of the used ion species and energy~\cite{Azarov2023}.
The former statement is confirmed by the observed conversion to the $\gamma$ phase for \hkl(010)-oriented substrates following \qty{140}{keV} Ne$^+$ irradiation with a fluence of \qty{3.5e16}{\ions\per\centi\meter\squared}(see Supplementary Note~1).

Increasing the fluence further to \qty{7.0e16}{\ions\per\centi\meter\squared} (\qty{52}{\dpa}) broadens the $\overline{8}$04 reflection asymmetrically towards a larger lattice plane spacing most likely due to strain accumulation.
After applying \qty{3.5e17}{\ions\per\centi\meter\squared}, which corresponds to \qty{260}{\dpa}, the $\gamma$\,444 reflection vanishes almost completely, likely due to significant disorder within the $\gamma$ layer. 

In the \ac{bfstem} image shown in Figure~\ref{fig:Structure}(c), the converted layer is visible with a thickness of \qty{\approx260}{nm} after \qty{140}{keV} Ne$^+$ irradiation with a fluence of \qty{3.5e16}{\ions\per\centi\meter\squared}. 
The \ac{fft} patterns shown in Figures~\ref{fig:Structure}(d) and (e) correspond to the regions marked in blue and red in Figure~\ref{fig:Structure}(c), respectively.
The \acp{fft} obtained from the different regions confirm that the top layer has the defective spinel cubic structure of \GGaO{}, while the unaffected substrate has the original monoclinic crystal structure of \BGaO{}.
As is evident from Figure~\ref{fig:Structure}(c), the $\gamma$ layer shows severe damage in particular close to the $\beta/\gamma$ interface. 
As can easily be observed in Figure~\ref{fig:Structure}(d), several diffraction spots in the \ac{fft} pattern of the cubic $\gamma$ phase, such as the \hkl{111} reflections (highlighted in blue circles), are blurry and suspected of spot broadening compared to the sharper \ac{fft} pattern of the monoclinic $\beta$ phase in Figure~\ref{fig:Structure}(e).
We attribute the spot broadening to crystal imperfections arising from the ion beam irradiation and the intrinsic vacancies present even in a perfect defective spinel structure as is the case here.

Different from the \ac{xrd} results presented above, we do observe weak and broad \hkl{111} reflections in the \ac{fft} pattern of the \ch{Ne+}-irradiated layer.
Careful analysis of the \ac{tem} samples using selected-area electron diffraction and convergent-beam electron diffraction confirms that due to the small number of \acp{apb} present in the limited volume of the \ac{tem} sample, the extinction of the \hkl{111} peaks is not perfect and they appear as broad and weak spots in the \ac{fft}.

\begin{figure}[btp]
    \centering
    \includegraphics[width=\linewidth]{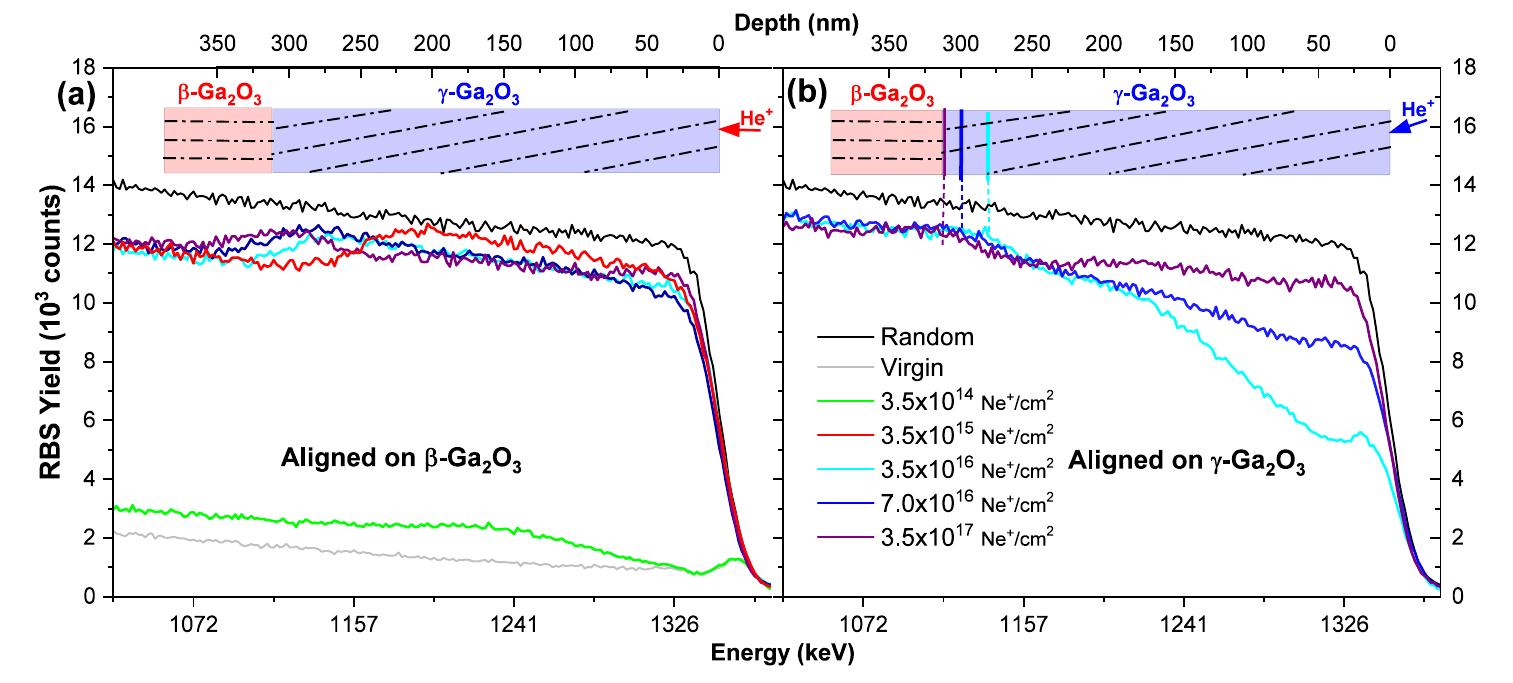}
    \caption{\ac{rbsc} spectra for \BGaO{} irradiated with different fluences. 
    The irradiated area is converted from \BGaO{} to \GGaO{} for \qty{3.5e16}{\ions\per\centi\meter\squared} and higher fluences.
    (a) \ac{rbsc} spectra obtained after aligning to the \hkl[-201] \BGaO{} channeling direction.
    (b) \ac{rbsc} spectra obtained after the phase transformation and realignment to the \hkl[111] \GGaO{} channeling direction. 
    The dashed lines in the sample configuration inset are only a sketch of the lattice planes and do not represent the actual misalignment.
    Of note, the depth scale is calculated for gallium atoms.}
    \label{fig:RBS}
\end{figure}

In Figure~\ref{fig:RBS}, \ac{rbsc} spectra are plotted for \GaO{} after \ch{Ne+} implantation of different fluencies to check the crystal quality after the irradiation.
Two sets of spectra are presented. 
%-------------------------------------------------------------------------------
%--------------------------------------------------------------------------
While the \ac{rbsc} spectra for \BGaO{} and \GGaO{} presented in Figure~\ref{fig:RBS}(a) were obtained with the ion beam parallel to the \BGaO{}--\hkl[-201] direction, the ion beam is aligned with the \GGaO{}--\hkl[111] direction for the spectra in Figure~\ref{fig:RBS}(b).
The obtained mismatch between the two is \qty{2.9(3)}{\degree} in $\chi$ (tilt angle).
This observation is consistent with previous \ac{ebsd} results presented in \cite{Azarov2023}, which report a slight misalignment between the otherwise fixed orientations of the $\beta$ substrate and $\gamma$ layer.
The sample irradiated with \qty{3.5e14}{\ions\per\centi\meter\squared} (light green) shows a slightly higher backscattering yield than the virgin (grey) sample due to the implantation-induced damage, but it still has good channeling properties and is far from the yield expected for an amorphous sample represented by the spectra obtained in a random orientation (black).
There is a clear increase of the backscattering yield upon irradiation with \qty{3.5e15}{\ions\per\centi\meter\squared}, shown in Figure~\ref{fig:RBS}(a).
At this fluence, the transformation is not finished (see Figure~\ref{fig:Structure}(a)) and the highly damaged mixed $\beta$/$\gamma$ layer has a high dechanneling rate.
As will be further discussed in the \ac{dbvepas} results, the high backscattering yield under this irradiation condition is attributed to the high concentration of defects.
Note that the yield drops again beyond a depth of \qty{200}{nm} which is deeper than the projected range of \qty{140}{keV} Ne$^+$ (approximately \qty{170}{nm}, calculated using SRIM~\cite{Ziegler2010}).
The drop indicates the transition from the irradiated/disordered $\beta$/$\gamma$ layer to the underlying ordered/unirradiated $\beta$ substrate.
Further increase of the fluence to \qty{3.5e16}{\ions\per\centi\meter\squared} and beyond does not result in an increase of the channeling yield in case the ion beam is aligned parallel to the \hkl[-201] direction of \BGaO{}, as illustrated in Figure~\ref{fig:RBS}(a). 
Note that \hkl[-201] direction is not the \hkl(-201) plane normal in \BGaO{} due to the non-orthogonal structure of the monoclinic lattice.
However, the previously observed drop in the yield shifts to greater depth (\qty{\approx285}{nm} at \qty{3.5e16}{\ions\per\centi\meter\squared}) with increasing fluence and corresponds well with the measurements performed in \ac{tem} (\qty{260}{nm}). 
Additionally, if we realign the beam to the channeling direction in \GGaO{}, as illustrated in Figure~\ref{fig:RBS}(b), the yield decreases significantly in the converted layer.
The \ac{rbsc} yield increases towards the lower energy part of the spectra and stays close to the random case after \qty{285}{nm}.
The latter fact is not surprising, as the beam is now misaligned with respect to the \BGaO{}--\hkl[-201] direction in the \BGaO{} bulk. 
Furthermore, the characteristic shape and reasonable crystallinity are maintained for an irradiation fluence of \qty{7e16}{\ions\per\centi\meter\squared}, corresponding to \qty{52}{\dpa}.
A significant increase of the \ac{rbsc} yield in the first \qty{150}{nm} is expected, since the maximum of the energy loss profile is at \qty{\approx115}{nm} for \qty{140}{keV} \ch{Ne+} ion implantation (see Supplementary Note~2 for SRIM calculation results), where the majority of the \ac{dpa} is generated.
Finally, after implantation of \qty{3.5e17}{\ions\per\centi\meter\squared}, corresponding to \qty{260}{\dpa}, the measurements result in a very high \ac{rbsc} yield, close to but not fully reaching the amorphous case.
Since the beam is aligned with the $\gamma$ layer, we assume that it has a high number of defects and \acp{apb} in the existing $\gamma$ phase, which consequently increases the \ac{rbsc} yield.
This is also evident from the \ac{xrd} results presented in Figure~\ref{fig:Structure}(b).

\begin{figure}[btp]
    \centering
    \includegraphics[width=0.785\linewidth]{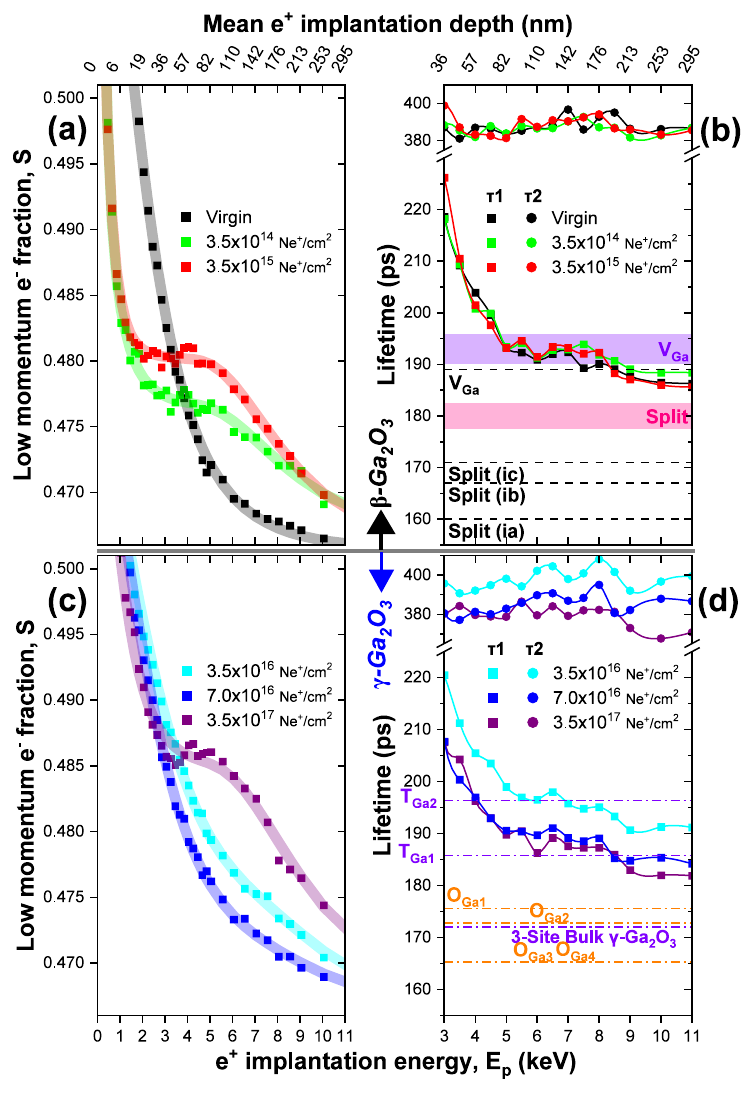}
    \caption{\ac{dbvepas} and \ac{vepals} results for Ne$^+$-irradiated samples with different fluences. 
    (a) and (c) show the S-parameter as a function of the positron implantation energy, 
    (b) and (d) show the positron lifetime as a function of the positron implantation energy.
    The colored transparent lines in (a) and (c) represent the fitted S-parameter curves generated from \ac{vepfit} for each sample.
    The dashed lines in (b) represent theoretically calculated lifetimes for various defect configurations in \BGaO{}~\cite{Karjalainen2020}, while the violet and pink bar correspond to lifetimes of the different defect configurations as interpreted by \emph{Tuomisto} for Fe-doped \BGaO{}~\cite{Tuomisto2023}.
    The dashed lines in (d) represent theoretically calculated lifetimes for various defect configurations in \GGaO{} (see table~\ref{tab:Lifetime}).
    }
    \label{fig:PALS}
\end{figure}

We performed \ac{dbvepas} and \ac{vepals} measurements (see Figure~\ref{fig:PALS}) to study the changes in defect type, structure, and concentration, for the $\beta$-to-$\gamma$ phase transition.
%--------------------------------------------------------------------
%-------------------------------------------------------------------
The detailed descriptions of these techniques are provided in the method section and elsewhere~\cite{Wagner2018}.
These methods utilize the effect that neutral and negatively charged vacancies, vacancy clusters, dislocations, or other open volume defects can trap positrons, where the relative contributions of valence and core electrons to the annihilation process varies, thus providing information about the local chemical environment (\ac{dbvepas}).
In addition, the lifetime of positrons in the material is affected by the local electron density (\ac{vepals}).
While \ac{dbvepas} provides combined information on both the identity and concentration of positron traps, the lifetime results obtained from \ac{vepals} offers complementary information by additionally quantifying the size of the defects.  

%%%%%%%%%%%%%%%%%%%%%%%%%%%%%%%%%%%%%%%
     %% DB-VEPAS BELOW %%
%%%%%%%%%%%%%%%%%%%%%%%%%%%%%%%%%%%%

In Figures~\ref{fig:PALS}(a) and (c), the S-parameter is plotted as a function of the positron implantation energy and mean positron implantation depth for the reference sample and various irradiation fluences used in this study~(see Supplementary Note~3 for S-W fractions).
After irradiation with \qty{3.5e14}{\ions\per\centi\meter\squared}, a \emph{plateau} in the S-parameter values is observed, indicating an increase of the defect density.
The S-parameter scales with the number of free volumes as long as positrons cannot easily diffuse due to a sufficient number of traps.
With a further increase to \qty{3.5e15}{\ions\per\centi\meter\squared}, a similar trend is observed, however with even greater S-parameter values in the \emph{plateau}.
When the defect concentration increases, there are more sites where positrons can get trapped and annihilate with low-momentum electrons, leading to an elevated S-parameter.
The observation gets supported by the calculation of the defect concentrations using \ac{vepfit} code~\cite{Saleh2013}, and the results are summarized in table~\ref{tab:Defect} (see Methods section for details on the calculation).
%----------------------------------------------------------------------
\begin{table}[tbp]
    \caption{\ac{vepfit} analysis results for \qty{140}{keV} Ne$^+$ irradiations of \BGaO{}.
    The defect concentration is calculated considering our \BGaO{} sample with the longest available effective diffusion length of \qty{52.4(5)}{\nm}.}
    \label{tab:Defect}
    \centering
  \begin{tabular}{@{}
  S[table-format=1.1e2]
  S[table-format=1.4(1), uncertainty-mode=compact]
  S[table-format=1.4(1), uncertainty-mode=compact]
  S[table-format=1.1(1)]
  S[table-format=3.0(2)]
  S[table-format=1.2e2, table-align-exponent=true]
  @{}}
    \toprule
    {Fluence} & {S-Parameter} & {W-Parameter} & {Diffusion length} & {Defective layer depth } & {Defect Concentration} \\ 
    {(Neon/cm$^2$)} & {(Layer)} & {(Layer)} & {L+ (nm)} & {d (nm)} & {C$_v$ (cm$^{-3}$)} \\
    \midrule
    0      & 0.4653(2) & 0.0490(1) & 27.7(2) & 0       & 1.81e18 \\
    3.5e14 & 0.4765(2) & 0.0446(1) & 8.3(5)  & 182(5)  & 2.7e19  \\
    3.5e15 & 0.4801(1) & 0.0449(1) & 3.0(2)  & 171(3)  & 2.1e20  \\
    3.5e16 & 0.4722(2) & 0.0485(1) & 39.4(4) & 325(14) & 4.24e17 \\
    7e16   & 0.4694(2) & 0.0500(1) & 32.8(3) & 326(30) & 8.56e17 \\
    3.5e17 & 0.4847(1) & 0.0468(1) & 9.6(2)  & 211(4)  & 1.59e19 \\
    \bottomrule
  \end{tabular}
\end{table}
%-----------------------------------------------------------------------
Note that the results obtained from the trapping of the positron is dominated by negatively charged defects and positively charged vacancies do not contribute to the effective defect concentration as measured by \ac{dbvepas}.
These positively charged vacancies---specifically, oxygen vacancies in our case---exhibit low positron trapping rates and small positron binding energies due to Coulomb repulsion. 
Moreover, their open volumes are too small to support a trapped state in simulations, therefore, oxygen vacancies do not trap positrons according to existing calculations~\cite{Karjalainen2020}.
Thus, by using the term \emph{defect concentration}, we in fact present the perspective observed by positrons---an effective cation vacancy density.

As shown in table~\ref{tab:Defect}, samples irradiated with \qtylist{3.5e14;3.5e15}{\ions\per\centi\meter\squared} have shorter positron diffusion length and higher defect concentration than the reference sample.
This is in line with the earlier described increased \ac{rbsc} yield (see fig.~\ref{fig:RBS}(a)) for these irradiation conditions (\qtylist{3.5e14;3.5e15}{\ions\per\centi\meter\squared}), which is an indication for an increased defect concentration. 
Interestingly, the previously observed \emph{plateau} in the S-parameter disappears after further increase of the irradiation fluence to \qty{3.5e16}{\ions\per\centi\meter\squared} (see fig.~\ref{fig:PALS}(c)).
From this, we can again calculate the defect concentration, which is \qty{4.24e17}{\per\centi\meter\squared}---three orders of magnitude lower as compared to the sample irradiated with only ${1}/{10}$ of the fluence.
This value, as is evident from table~\ref{tab:Defect}, is also lower than that of the reference sample.
However, from the above-presented \ac{xrd}, \ac{tem}, and \ac{rbsc} results we know that this material is not \BGaO{} anymore but \GGaO{}.
The decrease in defect density also explains the reduction in backscattering yield in the \ac{rbsc} signal within the first \qty{200}{nm} for the sample irradiated with \qty{3.5e16}{\ions\per\centi\meter\squared}, compared to the sample irradiated with \qty{3.5e15}{\ions\per\centi\meter\squared}, as shown in Figure~\ref{fig:RBS}(a).
Increasing the irradiation fluence by a factor of two to \qty{7e16}{\ions\per\centi\meter\squared} increases also the defect concentration by a factor of two (see table~\ref{tab:Defect}).
This is also evidenced from the steeper slope of the S-parameter depth profile, as plotted in Figure~\ref{fig:PALS}(c). 
A further increase to \qty{3.5e17}{\ions\per\centi\meter\squared} (\qty{260}{dpa}) causes the \emph{plateau} in the S-parameter data to reappear in the sub--\qty{200}{nm} range, as can be seen in Figure ~\ref{fig:PALS}(c).
As mentioned above, we attribute this to the accumulation of defects and an increase in their concentration.
The position of the \emph{plateau} also matches reasonably well with the prediction from SRIM for the maximum damage depth~(see Supplementary Note~2).
The high defect concentration for this sample goes hand in hand with the vanishing intensity of the \GGaO{} related peaks in the \ac{xrd} data presented in Figure~\ref{fig:Structure}(b) and the high \ac{rbsc} yield in Figure~\ref{fig:RBS}(b).

%%%%%%%%%%%%%%%%%%%%%%%%%%%%%%%%%%%%
     %% LIFETIME BELOW %%
%%%%%%%%%%%%%%%%%%%%%%%%%%%%%%%%%%%%
Positron lifetime results for Ne$^+$-irradiated samples are plotted in Figures~\ref{fig:PALS}(b) and (d).
Since the relative intensity of the positron signals are dominated by the short-lived positrons ($\tau_1$) rather than the long-lived positrons ($\tau_2$)(see Supplementary Note\,3), we will mainly focus on $\tau_1$ lifetimes indicating small-sized defects such as monovacancies or other simple atomic-scale defects~\cite{Rehberg1999}. 
Theoretically calculated lifetimes for specific defects in \BGaO{}, such as a gallium monovacancy (V$_{Ga}$) and a gallium split vacancy (Ga-divacancy with an interstitial Ga atom between two vicinal vacancies) are plotted as dashed lines in Figure~\ref{fig:PALS}(b) and correspond to \qty{189}{ps} and \qtyrange{160}{171}{ps},
respectively~\cite{Karjalainen2020}.
It should be noted that the approximations used by \emph{Karjalainen et al.}~\cite{Karjalainen2020} could lead to an underestimation of the lifetimes by not more than \qty{10}{\percent} due to the utilized  Boro\'nski-Nieminen local-density approximation~\cite{BoronskiPRB1986}.
Additionally, experimental lifetime measurements obtained on Fe-doped \BGaO{} crystals are plotted in Figure~\ref{fig:PALS}(b) at \qtyrange{190}{196}{ps}~(violet bar) and \qtyrange{177}{183}{ps}~(pink bar)~\cite{Karjalainen2021,Karjalainen2021a}.
Tuomisto assigned these lifetimes to the V$_{Ga}$ and gallium split vacancies, respectively~\cite{Tuomisto2023}.
Upon careful inspection of Figure~\ref{fig:PALS}(b), we see that the lifetime of the reference sample at \qtyrange{5}{8}{keV} positron implantation energy is slightly higher than the theoretically expected \qty{189}{ps} V$_{Ga}$ lifetime.
On the other hand, they do match nicely with the experimental values obtained for irradiated Fe-doped \BGaO{}~\cite{Karjalainen2021}.
These positron annihilation studies in the literature suggest that split vacancies are the dominant defect configuration in bulk \BGaO{}~\cite{Karjalainen2021}. 
The presence of V$_{Ga}$ defects in our case is most likely an artifact of the surface preparation and the resulting strain and damage (see Supplementary Note~3), contributing only insignificantly to the lifetimes reported in~\cite{Karjalainen2021}, which were measured with fast positrons instead of \ac{vepals}.
However, also in our low-energy near-surface analysis, we see that with increasing depth, the observed lifetimes get closer to the expected split vacancy lifetime values from reference~\cite{Karjalainen2021}.
Additionally, the ratio curves generated from the \ac{cdb} spectra confirm that the split-vacancy configuration is expected in the deeper regions of the reference sample, while V$_{Ga}$ defect configuration is observed in the near-surface region (see Supplementary Note~3), further validating our findings.

After irradiation with \qty{3.5e14}{\ions\per\centi\meter\squared} and \qty{3.5e15}{\ions\per\centi\meter\squared}, the samples show similar positron lifetimes, indicating similar defect types.
We therefore assume that in \BGaO{}, after irradiation with \qty{3.5e14}{\ions\per\centi\meter\squared} and \qty{3.5e15}{\ions\per\centi\meter\squared}, we find a large number of V$_{Ga}$ similar to those of the reference sample, but at a significantly higher concentration as deduced from the \ac{dbvepas} results above.
The measured positron $\tau_1$ lifetimes for these two samples are longer than the theoretically calculated \qty{189}{ps} for V$_{Ga}$.
It is plausible that the irradiated samples in the depth range of \qtyrange{142}{200}{nm} show, for example, a higher concentration of gallium divacancies (2~$\cdot$~V$_{Ga}$) as they are easier to stabilize within the crystal, or larger complexes of a gallium vacancy with oxygen vacancies (V$_{Ga}$ + n~$\cdot$~V$_O$, where $n > 1$), which increases the positron lifetime. 
Considering that theoretical models may underestimate positron lifetimes, it is possible that the observed disagreement is due to such an underestimation, and V$_{Ga}$ are indeed present in the irradiated samples.
The \ac{cdb} ratio curves further confirm this, indicating that the sample irradiated with \qty{3.5e15}{\ions\per\centi\meter\squared} exhibits a V$_{Ga}$ defect configuration near the surface region~(see Supplementary Note~3).

We also performed \ac{dft} calculations using the \ac{vasp} in order to obtain optimized bulk and defect configurations of the $\gamma$--phase.
In the first place, we estimated the bulk positron lifetimes of the $\gamma$--phase for 2, 3, and 4 site occupancy models~(see Supplementary Note~3).
These different site occupancy models are constructed using $1\times1\times3$ 160--atom cells and are idealized models from the literature~\cite{Ratcliff2022}.
The different $\gamma$--phase models exhibit different positron lifetimes because each model has a different localization of the positrons in the open volume~(see Supplementary Note~3). 
In addition, for the 3-site \GGaO{} model, which is the most energetically favorable structure for \GGaO{}~\cite{Ratcliff2022}, we estimated the bulk positron lifetime to be \qty{172}{ps}. 
Furthermore, we removed a gallium atom from either an octahedral (O$_{Ga1}$, O$_{Ga2}$, O$_{Ga3}$, O$_{Ga4}$) or tetrahedral (T$_{Ga1}$, T$_{Ga2}$) positions to simulate vacancy-induced positron trapping under non-equilibrium, irradiation-like conditions.
The resulting positron lifetimes for these V$_{Ga}$ configurations are listed in Table~\ref{tab:Lifetime}.
As already stated above, it is important to note that the theoretical models used here may underestimate the positron lifetimes. 
However, we will focus on the values obtained from our theoretical calculations and interpret the experimental results accordingly.

\begin{table}[tbp]
    \caption{Density functional calculations for the effect of different vacant ions on the positron lifetime.
    Ions at the tetrahedral (T$_{Ga}$) or octahedral (O$_{Ga}$) positions in the three-site \GGaO{} model have been considered. 
    }
    \label{tab:Lifetime}
    \centering
    \begin{tabular}{@{}lllS[table-format=3.0]@{}}
    \toprule
    \multirow{2}{*}{Ion} &
    \multicolumn{2}{c}{Nearest Neighbors (Ion-Distance [\AA])} &
    {Lifetime}\\
     & \multicolumn{1}{c}{Oxygen} & \multicolumn{1}{c}{Gallium} & {[ps]}\\
%    \begin{tabular}[c]{@{}c@{}}Ion\\        \end{tabular} &
%    \multicolumn{2}{c}{\begin{tabular}[c]{cc}\multicolumn{2}{c}{Nearest Neighbors (Ion-Distance [\AA])}\\ Oxygen & Gallium\end{tabular}} &
%    {\begin{tabular}[c]{@{}c@{}}Lifetime\\{[ps]}\end{tabular}} \\ 
    \midrule
    {Bulk}    &                                    &                        & 172 \\
    T$_{Ga1}$ & 1.83, 1.83, 1.88, 1.89             & 3.31, 3.33, 3.37, 3.40 & 186 \\
    T$_{Ga2}$ & 1.85, 1.85, 1.86, 1.86             & 3.07, 3.09, 3.23, 3.27 & 196 \\
    O$_{Ga1}$ & 1.93, 1.94, 1.98, 1.98, 2.06, 2.07 & 2.84, 2.94, 2.95, 3.16 & 176 \\
    O$_{Ga2}$ & 1.91, 1.92, 1.99, 2.04, 2.06, 2.07 & 2.87, 2.89, 2.94, 3.02 & 173 \\
    O$_{Ga3}$ & 1.94, 1.95, 1.96, 1.96, 2.05, 2.05 & 2.88, 2.89, 2.89, 2.89 & 165 \\
    O$_{Ga4}$ & 1.93, 1.94, 1.98, 1.98, 2.06, 2.07 & 2.84, 2.89, 2.89, 2.94 & 196 \\
    \bottomrule
    \end{tabular}
\end{table}

% Tga1 - O x 4 (1.83, 1.83, 1.88, 1.89)
%       Ga x 4 (3.31, 3.33, 3.37, 3.4)    

% Tga2 - O x 4 (1.85, 1.85, 1.86, 1.86)
%       Ga x 4 (3.07, 3.09, 3.23, 3.27)   

% Oga1 - O x 6 (1.93, 1.94, 1.98, 1.98, 2.06, 2.07)
%       Ga x 4 (2.84, 2.94, 2.95, 3.16) 

% Oga2 - O x 6 (1.91, 1.92, 1.99, 2.04, 2.06, 2.07)
%       Ga x 4 (2.87, 2.89, 2.94, 3.02)  

% Oga3 - O x 6 (1.94, 1.95, 1.96, 1.96, 2.05, 2.05)
%       Ga x 4 (2.88, 2.89, 2.89, 2.89)

% Oga4 - O x 6 (1.93, 1.94, 1.98, 1.98, 2.06, 2.07)
%       Ga x 4 (2.84, 2.89, 2.89, 2.94)     
%-----------------------------------------------------------------------

Even though the theoretical positron lifetime of the 3-site bulk \GGaO{} model (without added vacancies) is comparable to the split vacancy lifetimes of \BGaO{}, e.g. that of the split-i$_c$ configuration, the situation is different in the sense that the positron energy band is slightly dispersed as in the case of a free positron state (delocalized annihilation). 
We, therefore, interpret that the $\gamma$--phase can have a free positron state, but its lifetime can still be consistent with vacancies in the $\beta$--phase, since the positron mostly interacts with the vacant gallium sites in the $\gamma$--phase (see Supplementary Note 3).

%%%%%%%%%%%%%%%    3,5e16
In Figure~\ref{fig:PALS}(d), we plotted the calculated positron lifetimes for the 3-site $\gamma$ occupancy model with different gallium vacancy positions for the added vacancy.
Assuming the same trapping coefficient in \GGaO{} and \BGaO{}, the increased lifetime for the irradiation with \qty{3.5e16}{\ions\per\centi\meter\squared} (see Figure~\ref{fig:PALS}(d)) could be explained by the increased size of the positron traps in the $\gamma$--phase compare to the $\beta$--phase.
The measured positron lifetime of the sample irradiated with \qty{3.5e16}{\ions\per\centi\meter\squared} is in good agreement with the calculated lifetime for the 3-site $\gamma$ occupancy model when we create a V$_{Ga}$ on the T$_{Ga2}$ position (see table~\ref{tab:Lifetime}).
This suggests that after the ion-induced phase transition, the previously observed defect accumulation vanishes (compare Figure~\ref{fig:PALS}(a) with Figure~\ref{fig:PALS}(c)), and the $\gamma$--phase achieves its lowest energy configuration with additional V$_{Ga}$ at the T$_{Ga2}$ positions.
If a neon atom occupies isolated point defects (e.g., monovacancies), the positron lifetime would be strongly reduced, approaching values between those characteristic of V$_{Ga}$ and the bulk. 
However, we do not observe such lifetimes in our measurements.

%%%%%%%%%%%%%%%    7e16
Examining the positron lifetime results for the \qty{7e16}{\ions\per\centi\meter\squared} implantation in Figure~\ref{fig:PALS}(d), a significant reduction to \qty{\approx190}{ps} is observed, which is in between the \ac{dft}-obtained lifetimes for the T$_{Ga2}$ and T$_{Ga1}$ atomic configurations.
Our interpretation of this fact is that the $\gamma$--phase reaches a saturation for gallium vacancies at T$_{Ga2}$ positions, and the creation of gallium vacancies at T$_{Ga1}$ positions occurs by further increasing the irradiation fluence from \qty{3.5e16}{\ions\per\centi\meter\squared} to \qty{7e16}{\ions\per\centi\meter\squared}.
Therefore, the sample irradiated with \qty{7e16}{\ions\per\centi\meter\squared} likely contains defects in the form of gallium vacancies at both T$_{Ga2}$ and T$_{Ga1}$ positions. 
%%%%%%%%%%%%%%%    %3,5e17
An additional increase in fluence to \qty{3.5e17}{\ions\per\centi\meter\squared} leads to a further decrease in the lifetime, now matching well with the calculated lifetime for T$_{Ga1}$, as shown in Figure~\ref{fig:PALS}(d).
Consequently, \qty{3.5e17}{\ions\per\centi\meter\squared} irradiation conditions on \BGaO{}, corresponding \qty{260}{dpa}, create a \GGaO{} with a high concentration (see table~\ref{tab:Defect}) of gallium vacancies at the T$_{Ga1}$ position.
The average \ac{tde} for gallium vacancies at tetrahedral sites (\qty{22.9}{eV}) is higher than for the gallium vacancies at octahedral sites (\qty{20}{eV})~\cite{He2024}.
However, \ac{tde} is highly direction dependent in \GGaO{}, resulting in a higher \ac{tde} along the \hkl[111] lattice direction for gallium vacancies at octahedral sites than for tetrahedral sites~\cite{He2024}.
Based on the results presented in Figure~\ref{fig:PALS}(d), we can conclude that the formation of V$_{Ga}$ on octahedral sites is unlikely in our samples.
The increased defect concentration (see table~\ref{tab:Defect}), while showing similar relative intensities for $\tau_1$ and $\tau_2$ (see Supplementary Note~3), suggests that even if these defects are present, their detection is likely hindered by saturation trapping, with V$_{Ga}$ on tetrahedral sites being dominant.
It is worth noting that the implantation fluence of \qty{3.5e17}{\ions\per\centi\meter\squared} corresponds to a peak neon concentration of approximately 20~\text{at.\%} in the implanted region, therefore neon bubble formation is likely in this sample.
This is inline with the above observations that the $\gamma$ 444 peak disappears at this fluence (see Figure~\ref{fig:Structure}(b)) and the increased dechanneling in the RBS measurements (see Figure~\ref{fig:RBS}(b)).
In this case, implanted neon atoms accumulate in the voids, leading to the formation of bubbles, similar to the well known helium bubble formation~\cite{Allen2021, Gregor2016}. 
\Ac{pas} can reliably resolve defect clusters up to about 15 vacancies; for larger clusters, where the expected lifetimes are in the \qtyrange{350}{450}{ps} range, precise identification becomes more challenging~\cite{Cizek2018}.
The size of the neon bubbles formed at this fluence (\qtyrange{1}{5}{nm}) can be significantly larger and will not contribute the the observed positron lifetimes.
Smaller neon filled clusters are likely responsible for the observed small reduction of the $\tau_2$ lifetime for \qty{3.5e17}{\ions\per\centi\meter\squared} (see Figure~\ref{fig:PALS}(d)).

\begin{figure}[btp]
    \begin{center}
        \includegraphics[width=1\linewidth]{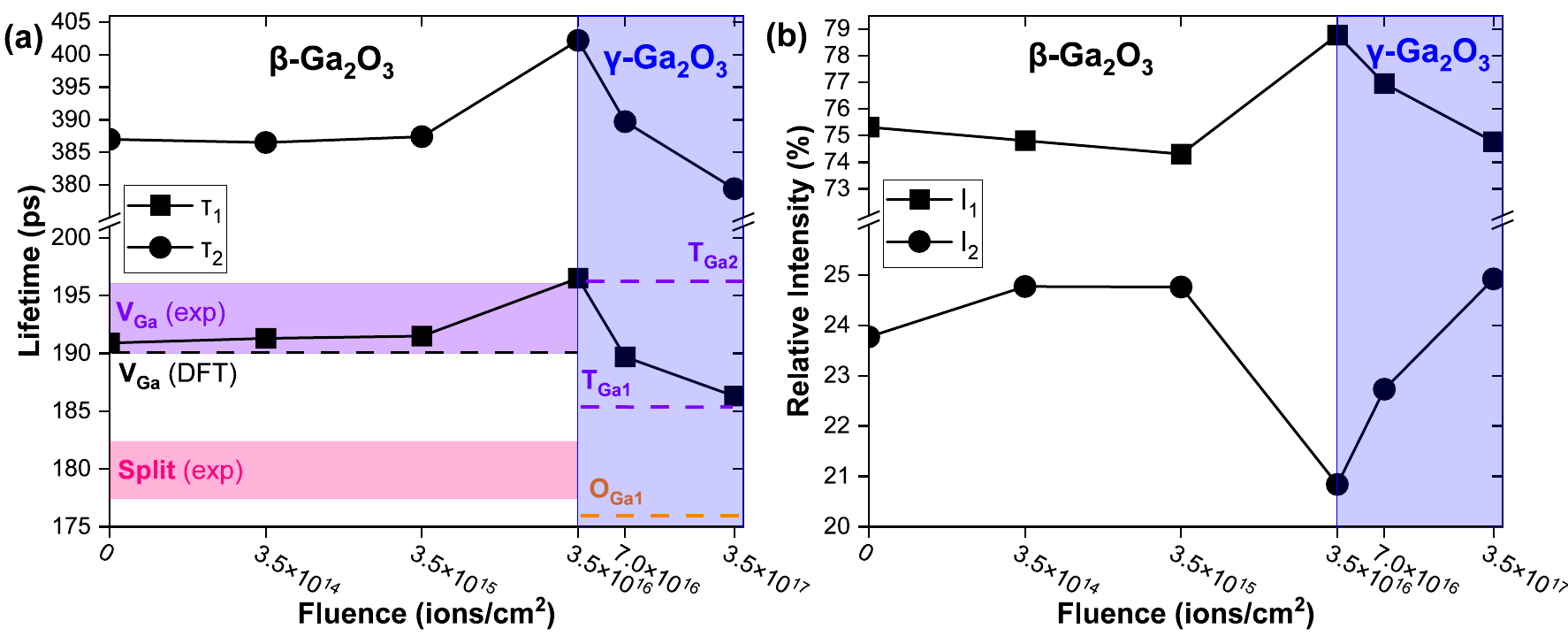}
    \end{center}
    \caption{\ac{vepals} lifetimes and relative intensities as a function of irradiation fluence obtained at a positron energy of \qty{6}{keV} or in depth of \qty{110}{nm}. 
    (a) Positron lifetimes of $\tau_1$ (fast annihilation pathway) and $\tau_2$ (slow annihilation pathway) as a function of irradiation fluence. 
    (b) Relative intensities of the short positron lifetime $\tau_1$ (I$_1$) and long positron lifetime $\tau_2$ (I$_2$) as a function of irradiation fluence.}
    \label{fig:ratiocurves}
\end{figure}

Figure~\ref{fig:ratiocurves} summarizes the evolution of the positron lifetimes and their relative intensities as a function of fluence.
As expected, little or no change is observed for $\tau_1$ and $\tau_2$ and their relative intensities before the phase transition from \BGaO{} to \GGaO{}.
Once the phase transition from \BGaO{} to \GGaO{} occurs between \qtylist{3.5e15; 3.5e16}{\ions\per\centi\meter\squared}, a distinct change is observed in all four monitored \ac{vepals} values, as shown in Figures~\ref{fig:ratiocurves}(a) and (b).
The change in lifetimes is expected, as the newly formed \GGaO{} has a defective spinel structure with more open volume as compared to the monoclinic crystal structure of \BGaO{}.
As shown above, the new $\tau_1$ lifetime corresponds to the calculated lifetimes of the T$_{Ga2}$ defects obtained from our \ac{dft} calculations.
However, also the relative intensities of $\tau_1$ and $\tau_2$ change, as shown in Figure~\ref{fig:ratiocurves}(b), indicating a relative increase of small-volume defects and a relative decrease of large-volume defects after the phase transition.
Increasing the irradiation fluence further leads to a decrease in both $\tau_1$ and $\tau_2$ lifetimes, with $\tau_1$ eventually approaching the \ac{dft}-calculated T$_{Ga1}$ lifetime value.
The observed changes in the relative intensities of the lifetime components suggest an increase in the proportion of vacancy clusters relative to single vacancies.
This is expected, as higher irradiation fluences promote the aggregation of single vacancies into larger defect structures.

In conclusion, we performed a multi-method analysis approach to investigate the defect structure of the \BGaO{} to \GGaO{} phase transition.
\ac{xrd} analysis of \hkl(-201)-oriented \BGaO{} following \qty{140}{keV} \ch{Ne+} irradiation indicates a phase transition occuring above a certain fluence threshold.
%Distinct $\gamma$\,222 and $\gamma$\,444 reflections were observed, while the expected $\gamma$\,111 and $\gamma$\,333 reflections were absent---likely due to the presence of a high density of \acp{apb} in the ion-induced $\gamma$ phase.
\ac{tem} imaging confirms the transition from the monoclinic to a defective spinel cubic crystal structure and reveals a transformed layer approximately \qty{260}{nm} thick.
Enhanced crystal quality was observed in the \ac{rbsc} spectra following the completion of the \GGaO{} phase transition.
\ac{dbvepas} and \ac{vepals} were employed to utilize the sensitivity of positrons as atomic-scale probes for non-destructive measurement and characterization of defects, such as single vacancies and their agglomerates.
\ac{vepfit} results show a reduced defect concentration and an increased positron diffusion length after irradiation with a fluence of \qty{3.5e16}{\ions\per\centi\meter\squared}, which serve as a clear signature of the ion-induced phase transition.
This reduced defect concentration is also in accordance with the enhanced crystal quality observed by \ac{rbsc}, after the phase transformation.
%The ratio curves generated from \ac{cdb} demonstrate that the sub-surfaces of differently oriented \BGaO{} substrates can be affected differently by wafer surface preparation, leading to different types of defects in the sub-surface region.
Positron lifetimes calculated for the three-site \GGaO{} phase model show an excellent agreement with the experimental results.
Notably, the reduction in positron lifetime with increasing irradiation fluence in \GGaO{} is an unexpected behavior for many other semiconductor materials and is a unique characteristic of \GGaO{}.
We attribute this to the increased formation of different gallium vacancy positions in the 3-site \GGaO{} model at high damage levels, which have a shorter positron lifetime.
Our results show that the relatively good crystalline quality as observed by \ac{xrd} and \ac{rbsc} in heavily irradiated \GaO{} is related to (a) the change in polymorph at around \qty{3.5e16}{\ions\per\centi\meter\squared} (in the case of \qty{140}{keV} \ch{Ne+} irradiation) and (b) the dominant formation of further defects in the already defective spinel structure of \GGaO{} at specific tetrahedral sites in the cubic lattice.
At high irradiation fluences, saturation of gallium vacancies at T$_{Ga2}$ sites and their aggregation into larger defect clusters are observed. 
Continued irradiation leads to the formation of single vacancies at T$_{Ga1}$ sites. 
This shift in vacancy formation within the $\gamma$ phase may explain and contribute to the high radiation tolerance of the $\beta$/$\gamma$ dual-phase polymorphic structure.

\section{Methods}\label{sec3}

Commercial \hkl(-201)-oriented \BGaO{} wafers from Novel Crystal Technology Inc. were irradiated with Ne$^+$ ions at an energy of \qty{140}{keV}. 
The applied fluence ranges from \qtyrange{3.5e14}{3.5e17}{\ions\per\centi\meter\squared} (see table~\ref{tab:Defect}).
The irradiations were performed at the Ion Beam Center of the Helmholtz-Zentrum Dresden\,--\,Rossendorf using a 500\,keV implanter (High Voltage Engineering Europa B.V., Model B8385).
Irradiations were performed using a 7\textdegree{} sample tilt to avoid channeling effects, and the maximum sample temperature during implantation was \qty{\approx60}{\celsius}.
All samples, measuring \qtyproduct{10x5}{mm}, were clamped at one corner and mounted on an aluminum plate for irradiation. 
Samples irradiated with fluences of \qty{3.5e14}{\ions\per\centi\meter\squared} and \qty{3.5e15}{\ions\per\centi\meter\squared} were exposed to an ion flux of \qty{1.41e12}{\ions\per\centi\meter\squared\per\second}. 
For higher fluences, a water-cooled mounting plate was used, and the ion flux was increased to approximately \qty{5.82e12}{\ions\per\centi\meter\squared\per\second}.

\Acf{xrd} measurements were carried out as $\theta/2\theta$ scans on a Rigaku SmartLab \qty{3}{kW} diffractometer with a Cu X-ray tube operated at \qty{40}{kV} and \qty{50}{mA}. 
A Ge \hkl(220) channel-cut monochromator limited the spectrum of the parallel beam to the K$_{\alpha1}$ line at \qty{0.15406}{nm}.
The sample was aligned in $\omega$ (incident angle) and $\chi$ (tilt angle) by its $\overline{2}01$ reflection of the \BGaO{} with the \hkl[102] direction parallel to the incident X-ray beam (at $\omega = 0$).

Cross-sectional \ac{tem} lamella preparation was carried out by in situ lift-out using a Thermo Fisher Helios 5 CX \ac{fib}-\acs*{sem} device. 
To protect the sample surface, a carbon cap layer was deposited beginning with electron-beam-induced and subsequently followed by Ga-\ac{fib}-induced precursor decomposition. 
Afterwards, the \ac{tem} lamella was prepared using a \qty{30}{keV} Ga-\ac{fib} with adapted currents. 
Its transfer to a 3-post copper lift-out grid (Omniprobe) was done with an EasyLift EX nanomanipulator (Thermo Fisher). 
To minimize sidewall damage, Ga ions with \qty{5}{keV} energy were used for final thinning of the \ac{tem} lamella to electron transparency. 
For comparison, a classical \ac{tem} cross-section, glued together in face-to-face geometry using G2 epoxy glue (Gatan), was prepared by sawing (Wire Saw WS 22, IBS GmbH), grinding (MetaServ 250, Bühler), polishing (Minimet 1000, Bühler), dimpling (Dimple Grinder 656, Gatan), and final Argon ion milling (Precision Ion Polishing System PIPS II 695, Gatan).
\ac{bfstem} imaging and \ac{cbed} (\ac{cbed} beam diameter about \qty{3}{nm}) were performed with a Talos F200X microscope (FEI) operated at an accelerating voltage of \qty{200}{kV}. 
\ac{hrtem} images were acquired with an image-C$_s$-corrected Titan 80-300 microscope (FEI) operated at \qty{300}{kV}. 
\ac{fft} anlaysis was done based on the recorded HR-TEM micrographs. 
Prior to (S)TEM analysis, the specimen mounted in a double-tilt low-background holder was placed for 8 seconds into a Model 1020 Plasma Cleaner (Fischione) to remove potential contamination. 

The used \ac{rbs} setup is connected to a \qty{2}{MV} Van de Graaff accelerator, and a \qty{1.7}{MeV} $^4$He$^+$ beam was used for all measurements.
The \ac{rbsc} measurements were performed along \hkl(-201) \BGaO{} and \hkl(111) \GGaO{} directions using a 170\textdegree{} backscattering geometry.
Notably, Ga parts of the \ac{rbsc} data were used in the analysis because of the significantly higher sensitivity of this method for the heavier Ga sublattice compared to the O sublattice.
Alignments for \ac{rbsc} to the \BGaO{} and \GGaO{} were performed using a \qty{\pm4}{\degree} rectangular frame scan.
For determining the channeling dips in \BGaO{} and \GGaO{}, the energy ranges of \qtyrange{0.99}{1.15}{MeV} and \qtyrange{1.24}{1.32}{MeV} were integrated, respectively. 
The minima obtained were used to find the planar channeling conditions, and the axial channeling conditions were subsequently calculated based on geometrical considerations.

\ac{dbvepas} measurements were conducted at the apparatus for in-situ defect analysis (AIDA)~\cite{Liedke2015} of the slow positron beamline (SPONSOR)~\cite{Anwand2012}. 
The kinetic energies of the positrons were adjusted to a discrete, monoenergetic values in the range of \Ep=\qtyrange[range-phrase=--,range-units=single]{0.05}{35}{keV} enabling their implantation into defined depths. 
The mean positron implantation depth $<z>$ was approximated using a simple-material-density ($\rho$) dependent formula~\cite{Dryzek2008}:
\begin{equation}
 <z>[nm]=\frac{36}{\rho[g\cdot cm^{-3}]}E_p^{1.62}[keV]
 \label{equ:mean-z}
\end{equation}

$<z>$ provides only a qualitative definition of the depth, while it does account for positron diffusion, however, is more accurate for materials with larger defect density.
In general, during implantation, positrons lose their kinetic energy due to thermalization and, after short diffusion, annihilate in delocalized lattice sites or localize in point defects and their agglomerations, emitting at least two anti-collinear \qty{511}{keV} gamma photons once they collide with electrons. 
The thermalized positrons have negligible momentum compared to the electrons, hence a broadening of the \qty{511}{keV} line is the consequence of the electron momentum. 
All the signals were measured with one or two high-purity Ge detectors (energy resolution of \qty{1.09+-0.01}{keV} or \qty{0.78+-0.02}{keV} at \qty{511}{keV} for single- and double-detector configuration, respectively). 
The broadening of the \qty{511}{keV} line is typically characterized by the two distinct parameters S and W defined as a fraction of the annihilation distribution in the middle (\qty{511+-0.84}{keV}) and outer regions (\qtyrange[range-phrase=--]{503.4}{508.12}{keV} and \qtyrange[range-phrase=--]{513.88}{518.61}{keV}), respectively.
The total area below the curve, which is utilized for the normalization of both parameters, is \qty{511+-16.24}{keV}.
The S-parameter is the fraction of positrons annihilating with low-momentum valence electrons and represents vacancy-type defects and their concentration. 
The W-parameter approximates overlap of positron wavefunction with high-momentum core electrons. 
Plotting the calculated S as a function of positron implantation energy, S(\Ep), provides depth-dependent information, whereas S-W plots are used to examine the atomic surrounding of the annihilation site and defect types~\cite{Clement1996}.

The \ac{cdb} measurements of the annihilation peak, where both annihilation photons are simultaneously recorded have been employed to investigate the atomic surrounding of the defect site. Typically, every chemical element has an unique shape of the cDB spectrum. Since thermalized positrons have negligible momentum compared to electrons, the effective momentum of annihilating electron-positron pair consist mostly of the electron momentum. As the consequence, measured Doppler shift in the energy of the annihilation-photons yields the momentum distribution of electrons that have annihilated positrons. In addition such a coincidence measurement suppresses the accidental events from the background by recording only events of the simultaneous detection of both annihilation-photons. The difference in energies of the two annihilation photons is $E_1$ - $E_2$ = 2 $\cdot$ $\Delta E = c\cdot p_L$, where $c$ is the speed of light and $p_L$ is the longitudinal component of the electron momentum to the direction of emitted annihilation photon \cite{Cizek2019}. cDB analysis has been conducted for the reference sample as well as for a characteristic ion fluence in order to discuss qualitative difference between the split and monovacancy defect configurations. The experimental findings have been compared with the theoretical considerations (see Supplementary Note\,3).

\Acf{vepals} measurements were conducted at the Mono-energetic Positron Source (MePS) beamline at HZDR, Germany~\cite{Wagner2018}. 
A CeBr$_3$ scintillator detector coupled to the Hamamatsu R13089-100 photomultiplier tube (PMT) was utilized for gamma photons detection.
The signals were processed by the SPDevices ADQ14DC-2X digitizer (14 bit vertical resolution and 2\,GS/s horizontal resolution)~\cite{Hirschmann2021}. 
The overall time resolution of the measurement system was better than \qty{0.250}{ns} and all spectra contained at least \qty{1e7}{counts}. 
A typical lifetime spectrum $N(t)$, the absolute value of the time derivative of the positron decay spectrum, is described by
\begin{equation}
N(t)=R(t)*\sum_{i=1}^{k+1} \frac{I_i}{\tau_i} e^\frac{-t}{\tau_i} + \text{"Background"},
\label{equ:pals-spectrum}
\end{equation}
where k is the number of different defect types contributing to the positron trapping, which are related to $k+1$ components in the spectra with the individual lifetimes $\tau_i$ and intensities $I_i$ ($\sum I_i=1$)~\cite{Rehberg1999}. 
The instrument resolution function $R(t)$ is a sum of two Gaussian functions with distinct intensities and relative shifts, both depending on the positron implantation energy, $E_p$.
It was determined by the measurement and analysis of a reference sample, i.e. single crystal yttria-stabilized zirconia, which exhibited a single well-known lifetime component. 
The background was negligible, hence fixed to zero.
All the spectra were deconvoluted using a non-linear least-squares fitting method, minimized by the Levenberg-Marquardt algorithm, employed within the fitting software package PALSfit~\cite{Olsen2007} into 2 major lifetime components, which directly evidence localized annihilation at 2 different defect types (sizes; $\tau_1$ and $\tau_2$).
Their relative intensities scale typically with the concentration of each defect type. 
In general, positron lifetime increases with defects size and open volume size. 
The positron lifetime and its intensity were probed as a function of positron implantation energy $E_p$ which was recalculated to the mean implantation depth $<z>$. 
The average positron lifetime $\tau_{av}$ is defined as $\tau_{av}$ = $\sum \tau_i I_i$.

Defect concentrations mentioned in the text and table~\ref{tab:Defect} were calculated with the formula:
\begin{equation}
 c_v=\frac{N}{\nu_v\tau_B}\left(\frac{L_{+,B}^2}{L^2_+}-1\right)
 \label{equ:concentration}
\end{equation}
where $N$ is the theoretical atomic density, $\nu_v$=\qty{1e15}{\per\second} is the positron trapping coefficient, $\tau_B$ is the measured positron lifetime, $L^2_{+,B}$=\qty{52.4\pm0.5}{nm} is the bulk positron diffusion length obtained from our best-quality reference sample, and $L^2_+$ is the effective positron diffusion length calculated from \ac{vepfit}. 
The atomic densities of \BGaO{} and \GGaO{} were estimated as  $N_{\beta}$=\qty{9.45e22}{at~cm^{-3}} and $N_{\gamma}$=\qty{9.54e22}{at~cm^{-3}}, respectively. $\tau_{B,\beta}$=\qty{135}{ps} and $\tau_{B,\gamma}$=\qty{173}{ps} for \BGaO{} and \GGaO{} were adopted from \cite{Karjalainen2020} and our \ac{dft} calculations.

The positron states and annihilations in the $\beta$-- and $\gamma$--phases of Ga$_2$O$_3$ were modeled using electronic structure methods and two-component theory for electron-positron systems \cite{BoronskiPRB1986}.
The results for the $\beta$--phase are obtained by projecting the three-dimensional momentum density data calculated in an earlier work \cite{Karjalainen2020}, and the $\gamma$--phase calculations were performed in this study.
Here, the \ac{vasp} code \cite{KresseCMS1996,KressePRB1996}, the projector augmented-wave (PAW) method \cite{BlochlPRB1994,KressePRB1999}, and the generalized gradient approximation for electron-electron exchange and correlation effects \cite{PerdewPRL1996} were used.
For the electron-positron correlation effects, the Boro\'nski-Nieminen local-density approximation was used \cite{BoronskiPRB1986}.
We assume that the localized positron does not influence the average electron density and take zero-positron-density limits of the correlation potential and enhancement factor.
The momentum density of annihilating pairs is calculated using the state-dependent model \cite{AlataloPRB1996}, reconstructed PAW wave functions for valence electrons \cite{MakkonenJPCS2005,MakkonenPRB2006}, and atomic orbitals for core electrons.
We perform orientational averaging of the Doppler spectra considering all possible orientations of the cubic $\gamma$ supercells.
Finally, before comparing with experiments, the Doppler spectra are convoluted with the experimental resolution function.
Our \ac{dft} calculations are based on the structural models reported by \emph{Ratcliff et al.}~\cite{Ratcliff2022}, which represent the most energetically stable configurations for each occupancy model in \GGaO{}. 
Given the intrinsic disorder of the Ga-sublattice, we consider three distinct models---2-site, 3-site, and 4-site---each corresponding to structures with two, three, and four occupied Ga sites, respectively.

\section*{Acknowledgment}

We acknowledge the M-ERA.NET Program for financial support via the GOFIB project supported by the tax funds on the basis of the budget passed by the Saxonian state parliament in Germany and administrated in Finland by the Research Council of Finland project number 352518. 
UB, GH, and NK acknowledge support by the COST Action CA19140 FIT4NANO.
This research was carried out at the Ion Beam Center and ELBE at the Helmholtz-Zentrum Dresden~--~Rossendorf~e.~V., a member of the Helmholtz Association. 
We would like to thank the facility staff for assistance.
We gratefully acknowledge Paul Chekhonin for his invaluable contributions to the \ac{cbed} analysis, and we extend our sincere thanks to Romy Aniol and Andreas Worbs for their expert assistance with \ac{tem} specimen preparation.
This work was partially supported by the Initiative and Networking Fund of the Helmholtz Association (FKZ VH-VI-442\,Memriox) and the Helmholtz Energy Materials Characterization Platform (03ET7015). 
We are grateful for CSC\,-Finnish IT Center for Science for generous computational resources.

\appendix

\bibliographystyle{MSP}
\bibliography{Defects-in-Ga2O3_v2}
\acresetall
\section{Supplementary Information}
\subsection{Supplementary Note 1 (Structure)}

\begin{figure}[!bp]
    \includegraphics[width=1\linewidth]{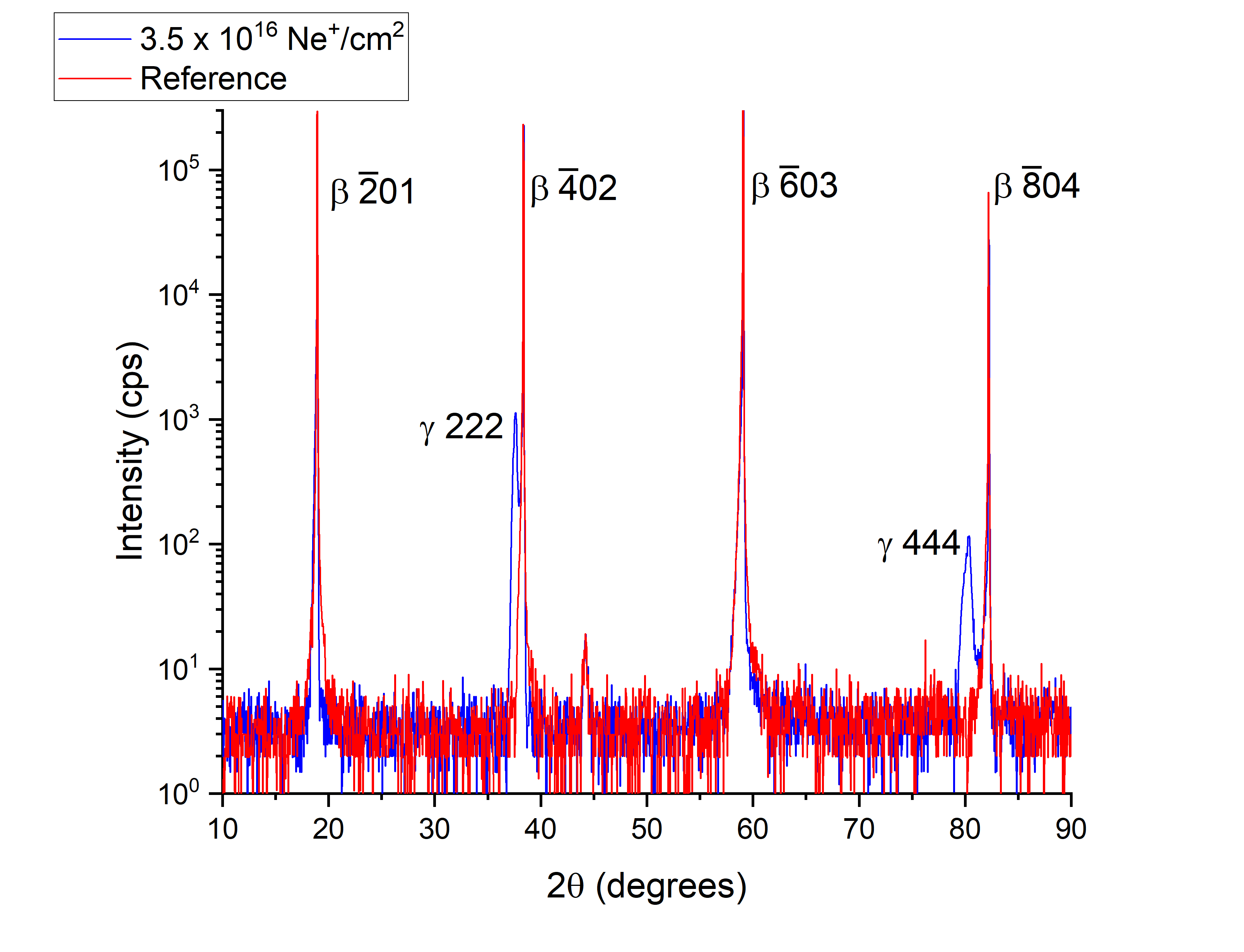}
    \caption{\ac{xrd} data for the ($\overline{2}01$)-oriented \BGaO{} virgin sample (red) and sample irradiated with \qty{140}{keV} Ne$^+$ using a fluence of \qty{3.5e16}{\ions\per\centi\meter\squared} (blue). While the $\gamma$\,222 and $\gamma$\,444 reflections are observed, the expected $\gamma$\,111 and $\gamma$\,333 reflections are missing.}
    \label{fig:S1}
\end{figure}

\begin{figure}[btp]
    \includegraphics[width=1\linewidth]{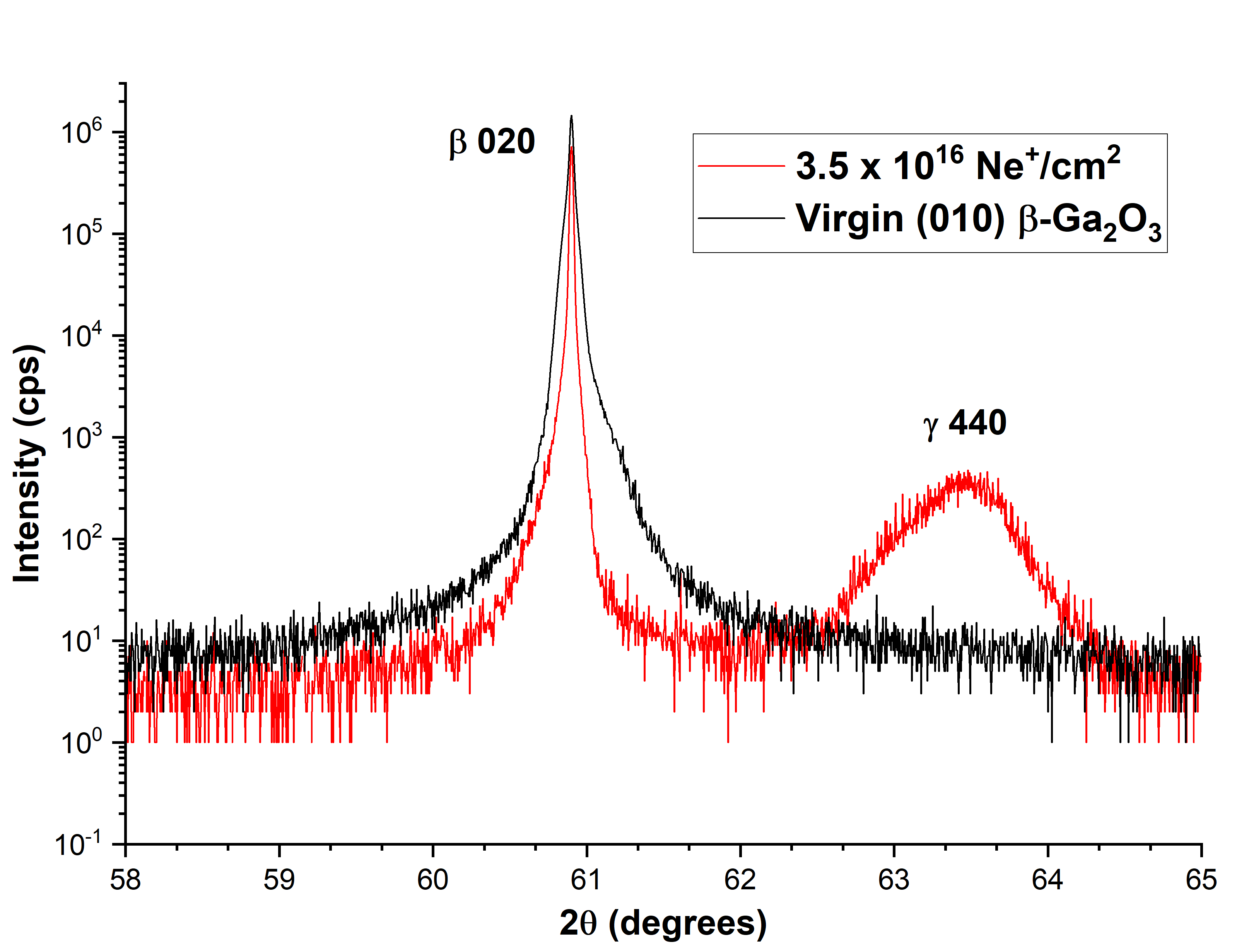}
    \caption{\ac{xrd} data for the \hkl(010)-oriented \BGaO{} virgin sample (black) and the \hkl(010)-oriented \BGaO{} sample irradiated with \qty{140}{keV} Ne$^+$ using a fluence of \qty{3.5e16}{\ions\per\centi\meter\squared} (red). 
    After the irradiation, the 440 reflection of \GGaO{} appeared.}
    \label{fig:S2}
\end{figure}

We performed \ac{xrd} analysis on a ($\overline{2}01$)-oriented \BGaO{} reference sample and a sample irradiated with \qty{140}{keV} Ne$^+$ at a fluence of \qty{3.5e16}{\ions\per\centi\meter\squared}. 
The $\theta/2\theta$ scans are shown in Figure~\ref{fig:S1}.
%-------------------------------------------------------------------
%-------------------------------------------------------------------
In the irradiated sample, new reflections appeared and were indexed as the 222 and 444 reflections of \GGaO{}. 
This phase transition was found to be independent of the surface orientation of the reference sample, as confirmed by Ne$^+$ irradiation under the same conditions on a \hkl(010)-oriented \BGaO{} sample, as shown in Figure~\ref{fig:S2}.
%-------------------------------------------------------------------
%-------------------------------------------------------------------
However, the allowed 111 and 333 reflections of the $\gamma$ phase, which should be observable, are absent in our measurements. 
We attribute this behavior to the presence of \ac{apb} within the $\gamma$ layer. 
\acp{apb} are planar defects commonly observed in $\gamma$--\ch{Al2O3}~\citeS{Rudolph2019}, typically formed as a result of dislocation movements. 
\emph{Tang et al.} reported a significant presence of \acp{apb} in \GGaO{} films deposited on \ch{MgAl2O4} substrates using metal-organic chemical vapor deposition~\cite{Tang2024}.
\acp{apb} can also form as a result of interactions between different domains within the material. 
In ion-induced \GGaO{}, localized nucleation and growth of the $\gamma$ phase after ion irradiation can lead to the formation of separate domains.
When two domains of \GGaO{} with different atomic arrangements meet, \acp{apb} can form.  Similarly, \emph{García-Fernández et al.} demonstrated the formation and presence of \acp{apb} in ion-irradiated \GGaO{}~\cite{GarciaFernandez2024}.  

\begin{figure}[btp]
    \includegraphics[width=1\linewidth]{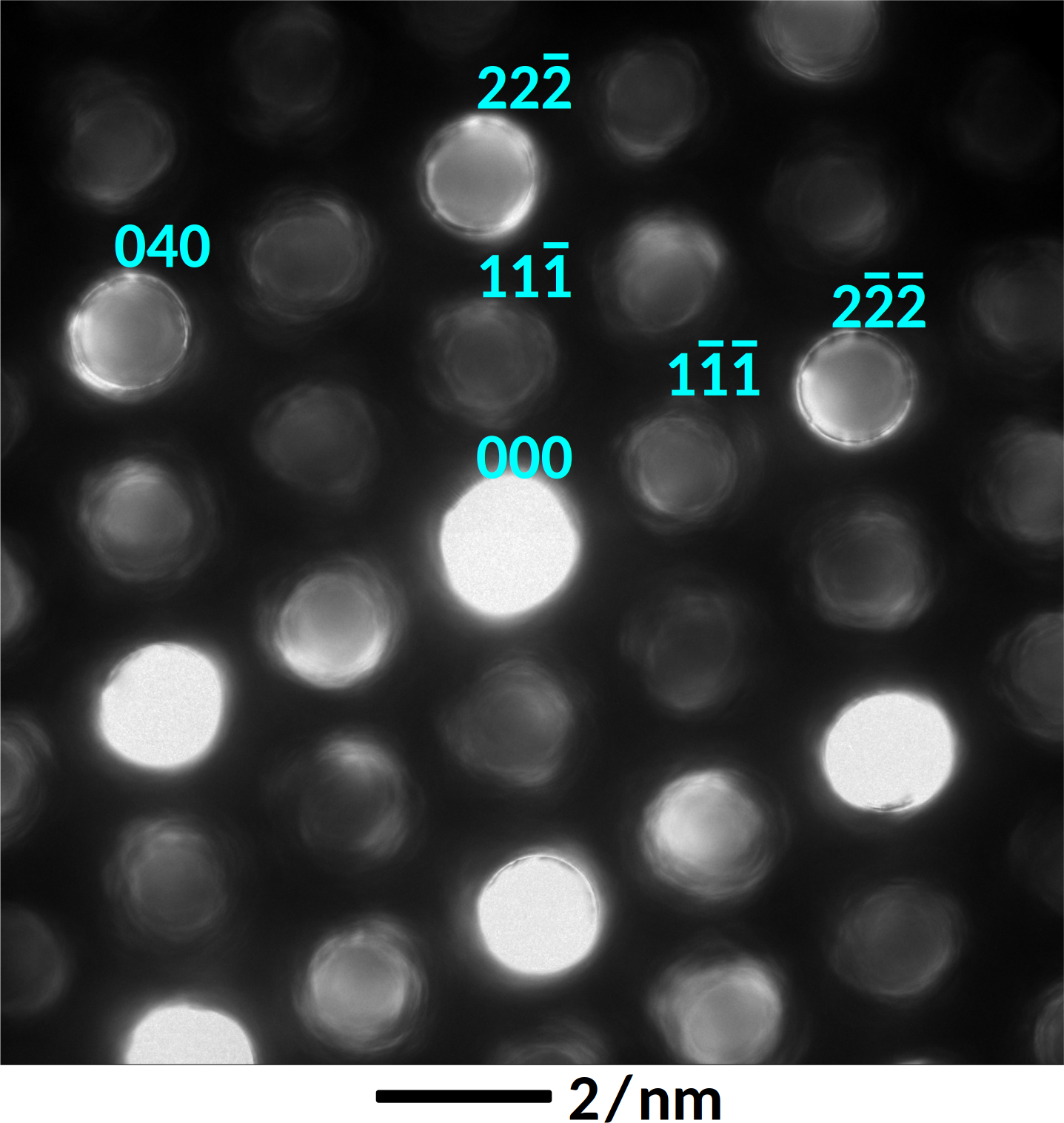}
    \caption{\ac{cbed} pattern from ion-induced (\qty{140}{keV} Ne$^+$ with a fluence of \qty{3.5e16}{\ions\per\centi\meter\squared}) \GGaO{} layer in \hkl[101] zone axis. 
    The convergent beam has a spot size of \qty{3}{nm}.}
    \label{fig:S3}
\end{figure}

In the present study, we have identified clear evidence of \acp{apb} in \GGaO{} samples by performing \ac{cbed}.
As shown in Figure~\ref{fig:S3}, the reflections in the observed \ac{cbed} pattern recorded in the \GGaO{} layer are not clear discs, but appear to be an overlap of several displaced discs.
%-------------------------------------------------------------------
%-------------------------------------------------------------------
This is particularly evident in $\gamma$\,111 and other low-intensity reflections. 
The explanation for this appearance is given by the shape of the reciprocal rods in the presence of additional boundaries, such as \acp{apb}.
In the \ac{cbed} experiment, the transmitted \GGaO{} layer has a thickness (along the \ac{tem} electron beam direction) of about \qty{100}{nm}.
If the average spacing of the \acp{apb} is << \qty{100}{nm}, the electron beam is likely to encounter several \acp{apb} along its path, leading to the observed diffraction disc splitting. 
This is consistent with the observations of \emph{García-Fernández et al.} where a similar ion-induced \GGaO{} layer appears to consist of \GGaO{} domains separated by \acp{apb} with a spacing of less than \qty{10}{nm}~\cite{GarciaFernandez2024}.

\ac{apb} can induce phase shifts within the crystal lattice which lead to destructive interference for specific reflections.
As shown in Table~\ref{tab:S1}, we calculated the phase shifts in different directions for the 111, 222, 333, and 444 reflections using the formula

                  \[ e^{2\pi i(hu + kv + lw)} \]

where \hkl(hkl) denotes the Miller indices, and \hkl[uvw] is the shift.
%-------------------------------------------------------------
\begin{table}
\caption{Phase shifts in different crystal lattice directions by \(\frac{1}{4}\) are calculated for the $\gamma$\,111, $\gamma$\,222, $\gamma$\,333, and $\gamma$\,444 reflections. 
Odd multiples of $\pi$ cause destructive interference, while even multiples of $\pi$ cause constructive interference.  
    }
    \label{tab:S1}
\centering
%\begin{tabular}{c|cccc|}
%\cline{2-5}
%                                  & \multicolumn{4}{c|}{\textbf{Reflections}} \\ \hline
%\multicolumn{1}{|c|}{\textbf{Shift Direction}} & \multicolumn{1}{c|}{111} & \multicolumn{1}{c|}{222} & \multicolumn{1}{c|}{333} & 444 \\ \hline
%\multicolumn{1}{|c|}{{[}110{]}}   & $\pi$  & 2$\pi$  & 3$\pi$  & 4$\pi$       \\ \cline{1-1}
%\multicolumn{1}{|c|}{{[}101{]}}   & $\pi$  & 2$\pi$  & 3$\pi$  & 4$\pi$       \\ \cline{1-1}
%\multicolumn{1}{|c|}{{[}011{]}}   & $\pi$  & 2$\pi$  & 3$\pi$  & 4$\pi$       \\ \cline{1-1}
%\multicolumn{1}{|c|}{{[}-110{]}}  & 0      & 0       & 0       & 0            \\ \cline{1-1}
%\multicolumn{1}{|c|}{{[}-10-1{]}} & -$\pi$ & -2$\pi$ & -3$\pi$ & -4$\pi$      \\ \cline{1-1}
%\multicolumn{1}{|c|}{{[}-1-10{]}} & -$\pi$ & -2$\pi$ & -3$\pi$ & -4$\pi$      \\ \cline{1-1}
%\multicolumn{1}{|c|}{{[}-101{]}}  & 0      & 0       & 0       & 0       \\ \hline
%\end{tabular}
\begin{tabular}{ccccc}
\toprule
\textbf{Shift}  & \multicolumn{4}{c}{\textbf{Reflections}} \\ 
\textbf{Direction} & 111 & 222 & 333 & 444 \\ 
 \midrule
{\hkl[110]}   & $\pi$  & 2$\pi$  & 3$\pi$  & 4$\pi$       \\ 
{\hkl[101]}   & $\pi$  & 2$\pi$  & 3$\pi$  & 4$\pi$       \\ 
{\hkl[011]}   & $\pi$  & 2$\pi$  & 3$\pi$  & 4$\pi$       \\ 
{\hkl[-110]}  & 0      & 0       & 0       & 0            \\ 
{\hkl[-10-1]} & $-\pi$ & $-2\pi$ & $-3\pi$ & $-4\pi$      \\ 
{\hkl[-1-10]} & $-\pi$ & $-2\pi$ & $-3\pi$ & $-4\pi$      \\ 
{\hkl[-101]}  & 0      & 0       & 0       & 0            \\ 
\bottomrule
\end{tabular}
\end{table}
%-------------------------------------------------------------
When the lattice is shifted by \(\frac{1}{4}\)  along specific directions, such as \hkl[110], \hkl[101], and \hkl[011], a resulting phase shift of $\pi$ leads to destructive interference for the $\gamma$\,111 and $\gamma$\,333 reflections. 
The splitting of \ac{cbed} peaks in multiple directions in Figure~\ref{fig:S3} indicates that the shift of the crystal lattice occurs along several directions, rather than a single one. 
Consequently, the absence of $\gamma$\,111 and $\gamma$\,333 reflections in \ac{xrd} can be attributed to the high density of \acp{apb} that are present in the ion-induced \GGaO{} layer. 

The observed broadening and weakening of these reflections in \ac{cbed} patterns and \ac{fft} patterns in Figure~\ref{fig:Structure}(d) of the main manuscript, in contrast to their absence in \ac{xrd}, can be explained by differences in the interaction/coherence volumes of electrons and X-rays within the samples. 
The electron interaction depth ($\approx \qty{50}{nm}$) is significantly smaller than the X-ray interaction depth (average \qty{5}{\micro\metre}). 
Since the X-ray interaction depth comprises the whole $\gamma$ layer, the volumes of domains separated by \acp{apb} are equal, leading to nearly perfect extinction for the $\gamma$\,111 and $\gamma$\,333 reflections in \ac{xrd}. 
In contrast, the smaller interaction volume of electrons used in \ac{tem} (\ac{fft} imaging and \ac{cbed}) does not comprise equal volume fractions of the two domain types. 
Therefore, the $\gamma$\,111 and $\gamma$\,333 reflections are not completely extinct.

\begin{figure}[btp]
    \includegraphics[width=1\linewidth]{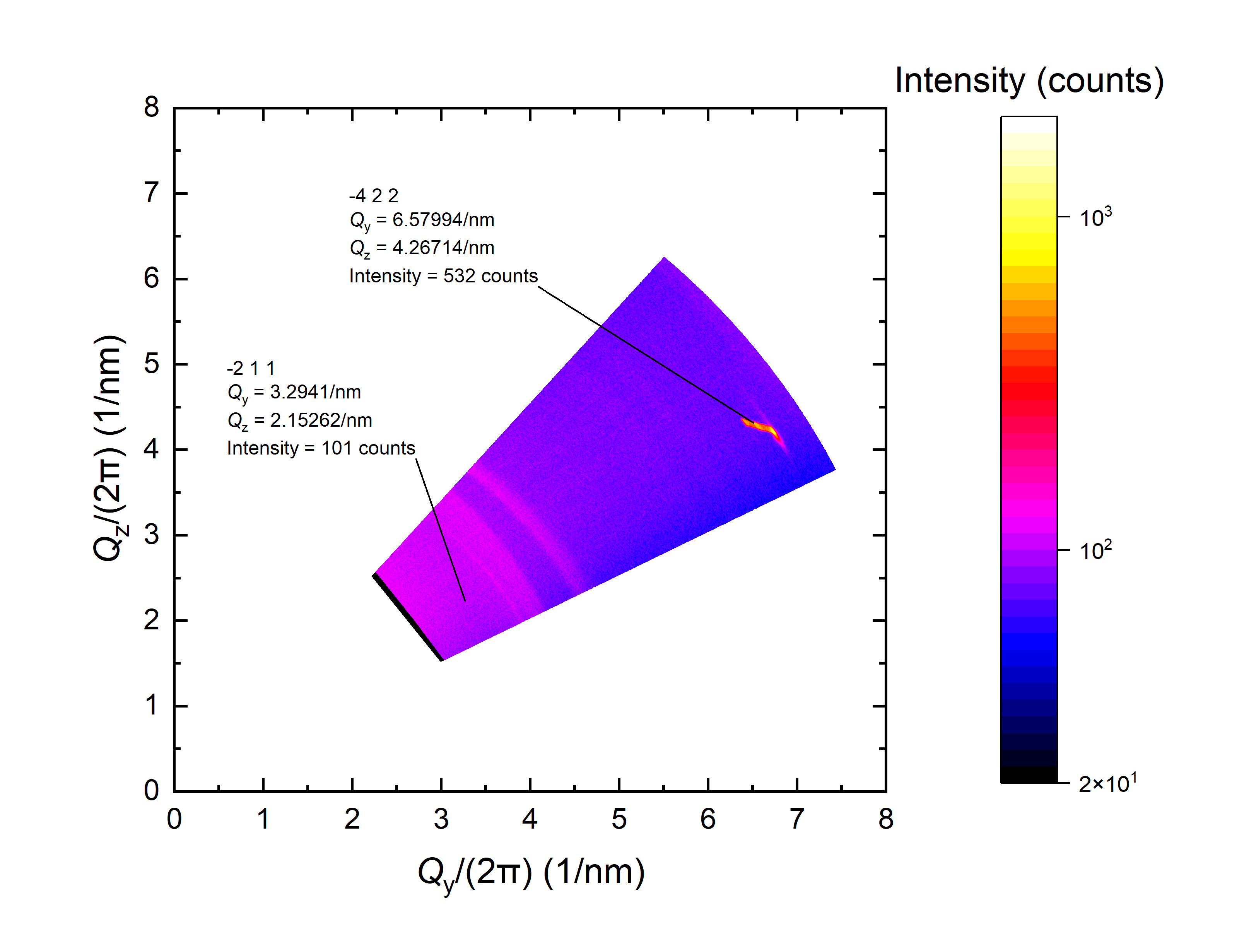}
    \caption{\ac{rsm} measurement of the \BGaO{} reference substrate. 
    The expected $\beta$\,$\overline{4}22$ reflection is observed, while the suspected forbidden $\beta$\,$\overline{2}11$ reflection is absent.}
    \label{fig:S4}
\end{figure}

During the investigation of the \ac{tem} lamella prepared by lift-out using the \ac{fib} technique (employing Ga$^+$), we observed broadened and weak spots in the \BGaO{} reference sample along the \hkl[102] zone axis. 
Initially, we suspected poor substrate quality as a potential cause. 
However, high substrate quality was confirmed through \ac{rsm} measurements, as shown in Figure~\ref{fig:S4}.
%-------------------------------------------------------------------
%-------------------------------------------------------------------
We then hypothesized that the final Ga$^+$ \ac{fib} polishing during \ac{tem} lamella preparation might induce unexpected interactions with the \BGaO{} substrate.

\begin{figure}[btp]
\centering
    \includegraphics[width=0.75\linewidth]{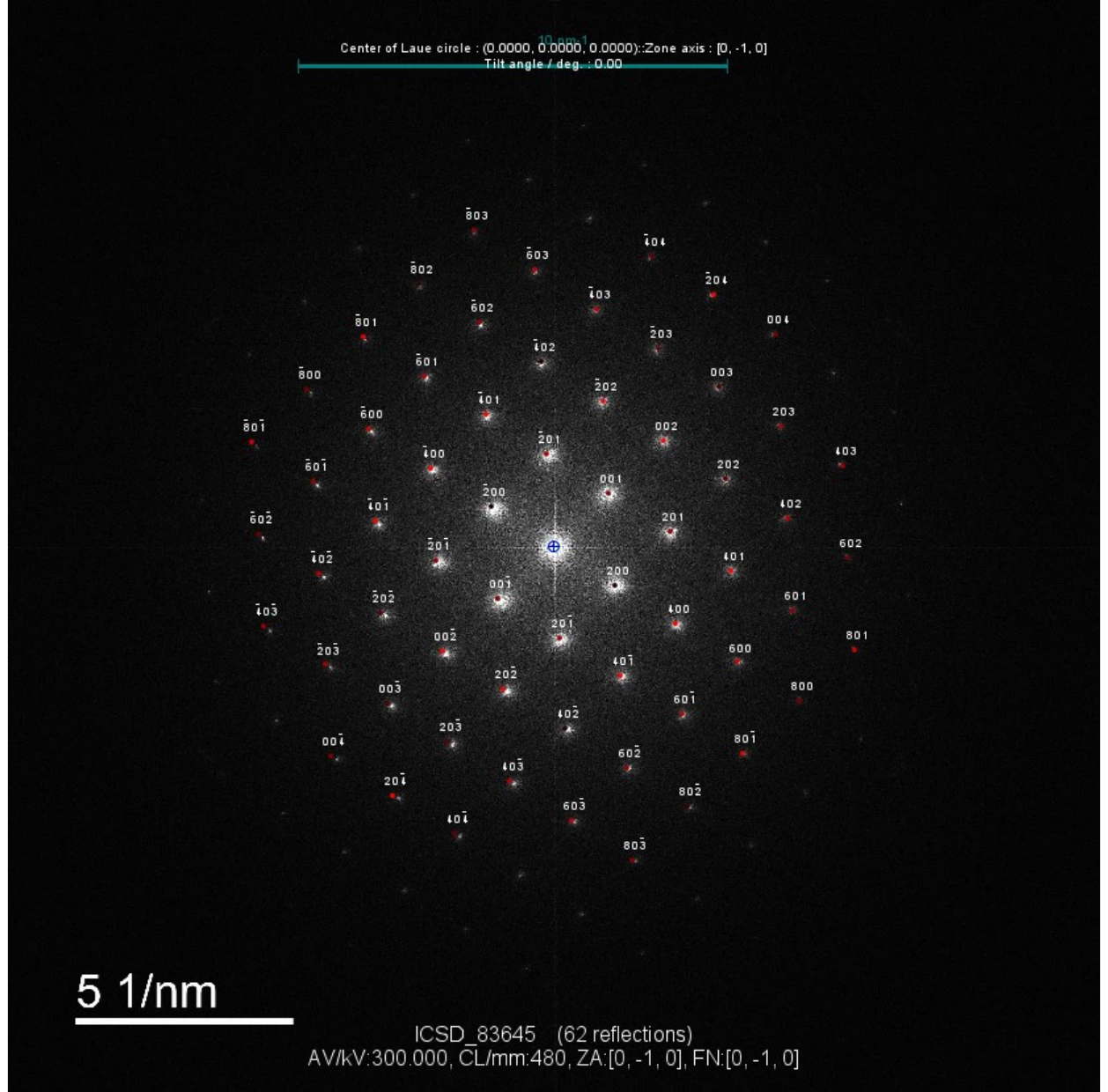}
    \caption{\ac{fft} pattern of a \ac{hrtem} image from \BGaO{} in \hkl[0-10] zone axis.
    The lamella was prepared using final Argon ion polishing (Lamella 1).}
    \label{fig:S5}
\end{figure}

\begin{figure}[btp]
\centering
    \includegraphics[width=0.5\linewidth]{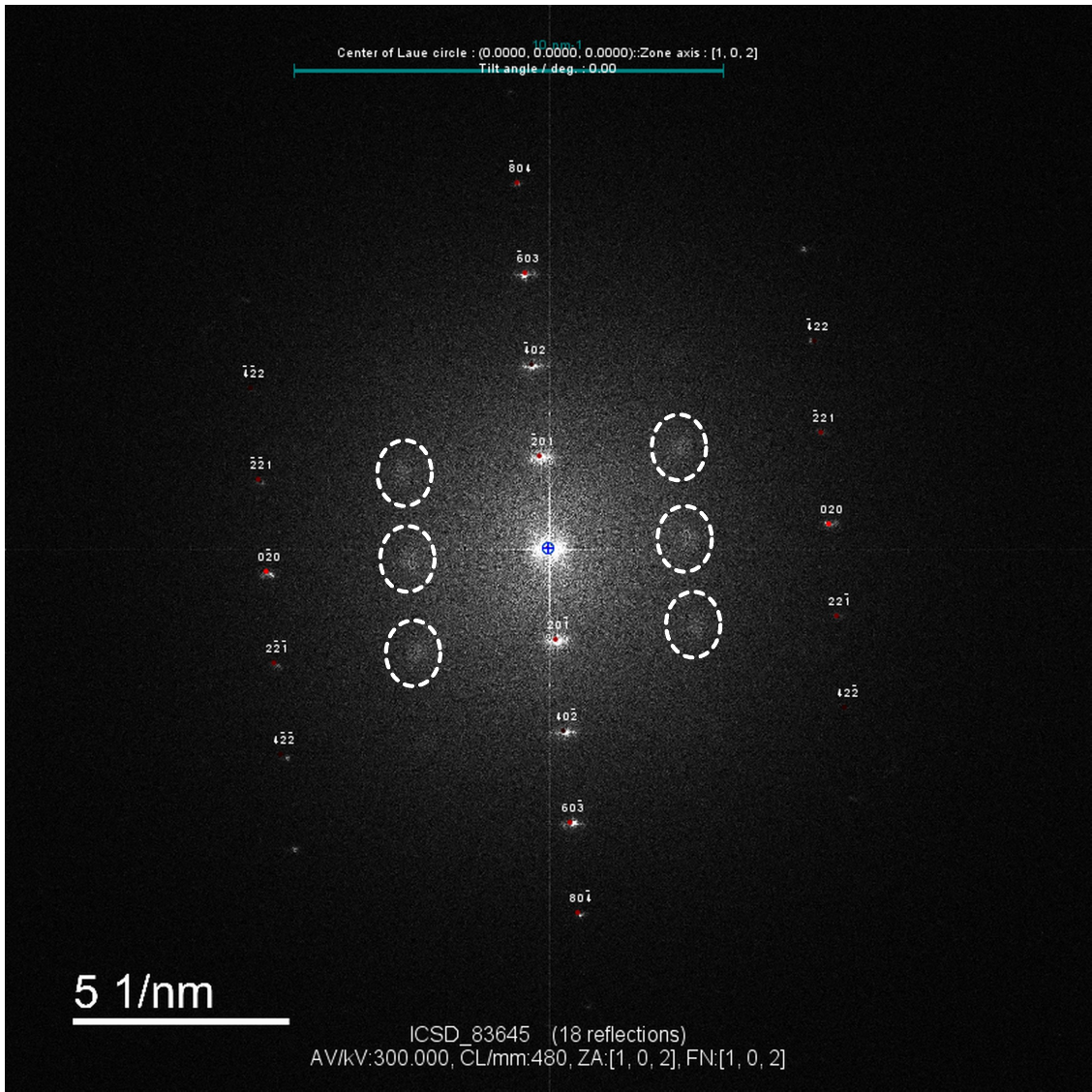}
    \caption{ac{fft} pattern of a \ac{hrtem} image from \BGaO{} in \hkl[102] zone axis.
    The lamella was prepared using final Argon ion polishing.
    Broadened and weak spots (highlighted with white-doted circles) are observed, which do not correspond to \BGaO{} reflections (Lamella 2).}
    \label{fig:S6}
\end{figure}

To test this, we prepared two classical \ac{tem} specimens employing final argon ion milling.
In particular, ($\overline{2}01$)-oriented \BGaO{} treated with \qty{140}{keV} \ch{Ne+} with a fluence of \qty{3.5e16}{\ions\per\centi\meter\squared} was cut in \hkl[0-10] and \hkl[102] in-plane zone axis orientations. 
The corresponding \ac{fft} patterns of \ac{hrtem} images are shown in Figures~\ref{fig:S5} and~\ref{fig:S6}, respectively.
%-------------------------------------------------------------------
%-------------------------------------------------------------------
%-------------------------------------------------------------------
%-------------------------------------------------------------------
Interestingly, only the \hkl[102] zone axis pattern reveals the unexpected reflections (highlighted with white-dotted circles). 

\begin{figure}[btp]
\centering
    \includegraphics[width=0.75\linewidth]{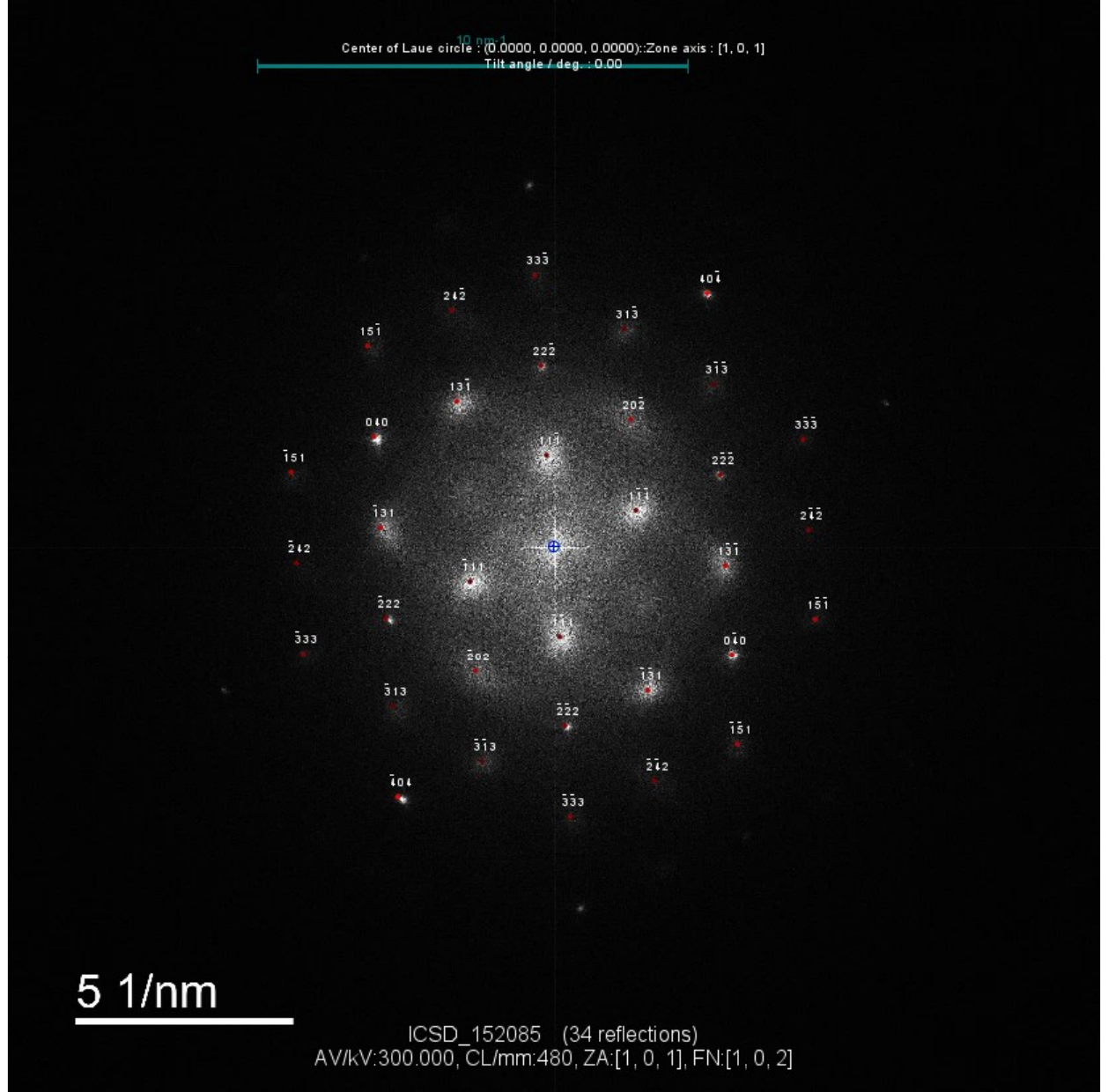}
    \caption{\ac{fft} pattern of a \ac{hrtem} image from ion-induced (\qty{140}{keV} Ne$^+$ with a fluence of \qty{3.5e16}{\ions\per\centi\meter\squared}) \GGaO{} layer in \hkl[101] zone axis. 
    The expected $\gamma$\,111 reflections are broad and weak (shown with blue circles) (Lamella 1).}
    \label{fig:S7}
\end{figure}

\begin{figure}[btp]
\centering
    \includegraphics[width=0.5\linewidth]{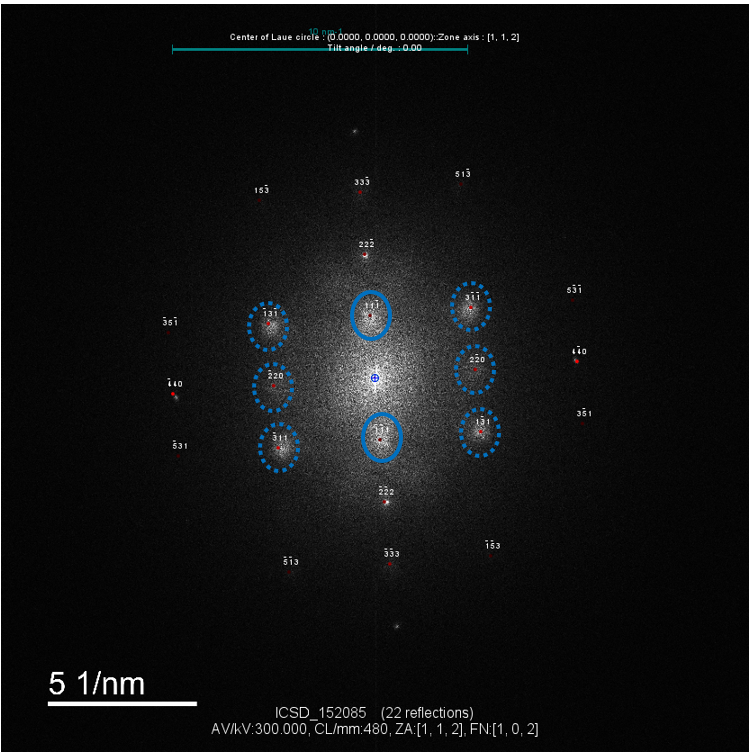}
    \caption{\ac{fft} pattern of a \ac{hrtem} image from ion-induced (\qty{140}{keV} Ne$^+$ with a fluence of \qty{3.5e16}{\ions\per\centi\meter\squared}) \GGaO{} layer in \hkl[112] zone axis.
    Expected $\gamma$\,111 reflections (shown with blue circles) and some others (shown with blue-dotted circles) are broad and weak (Lamella 2).}
    \label{fig:S8}
\end{figure}

\begin{figure}[btp]
\centering
    \includegraphics[width=0.5\linewidth]{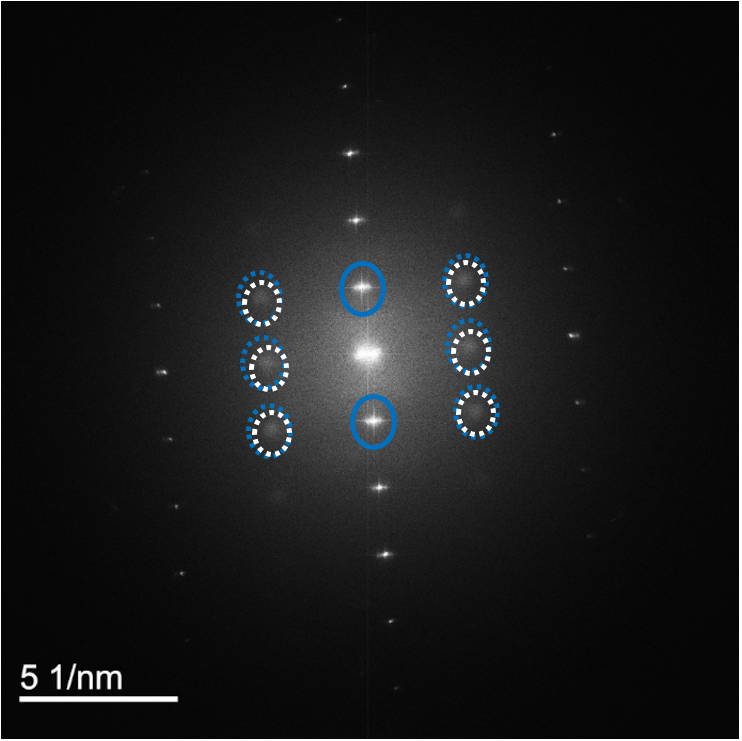}
    \caption{Two superimposed \acp{fft} from substrate \BGaO{} in \hkl[102] zone axis (figure~\ref{fig:S6}) and ion-induced (\qty{140}{keV} Ne$^+$ with a fluence of \qty{3.5e16}{\ions\per\centi\meter\squared}) \GGaO{} layer in \hkl[112] zone axis (figure~\ref{fig:S8}).}
    \label{fig:S9}
\end{figure}

Additionally, Figures~\ref{fig:S7} and ~\ref{fig:S8} show \acp{fft} of an ion-induced (\qty{140}{keV} Ne$^+$ with a fluence of \qty{3.5e16}{\ions\per\centi\meter\squared}) \GGaO{} layer along two different zone axes.
%-------------------------------------------------------------------
%-------------------------------------------------------------------
%-------------------------------------------------------------------
%-------------------------------------------------------------------
Expected $\gamma$\,111 reflections (highlighted with blue circles) and other reflections (highlighted with blue-dotted circles) appear broadened and weak. 
Interestingly, when the FFTs are superimposed (Figure~\ref{fig:S9}), the unexpected blurry reflections from Figure~\ref{fig:S6} matched the reflections characteristic of the $\gamma$\,phase (Figure~\ref{fig:S8}).
%-------------------------------------------------------------------
%-------------------------------------------------------------------
Thus, we hypothesize that the lamella preparation process causes, or at least starts, a phase transition from the $\beta$ to the $\gamma$\,phase, which has a high amount of antiphase boundaries. 
The reason this phenomenon is observed only along the [102] zone axis and not the \hkl[0-10] zone axis remains an open question.
However, as discussed later in Supplementary Note~3, defects exhibit different behaviors in the same material depending on the surface orientation. 
The impact of lamella preparation may vary between different zone axes, potentially due to differences in planar atomic density.

\subsection{Supplementary Note 2 (dpa)}

The \ac{dpa} is a measure of irradiation damage.
It quantifies how often an atom in the crystal is displaced from its lattice position. Conventionally, \ac{dpa} can be calculated via;

\[
\text{dpa} = \frac{N_\text{vac}^\text{max} \cdot \text{Fluence}}{n_a}
\]

\begin{figure}[btp]
    \includegraphics[width=1\linewidth]{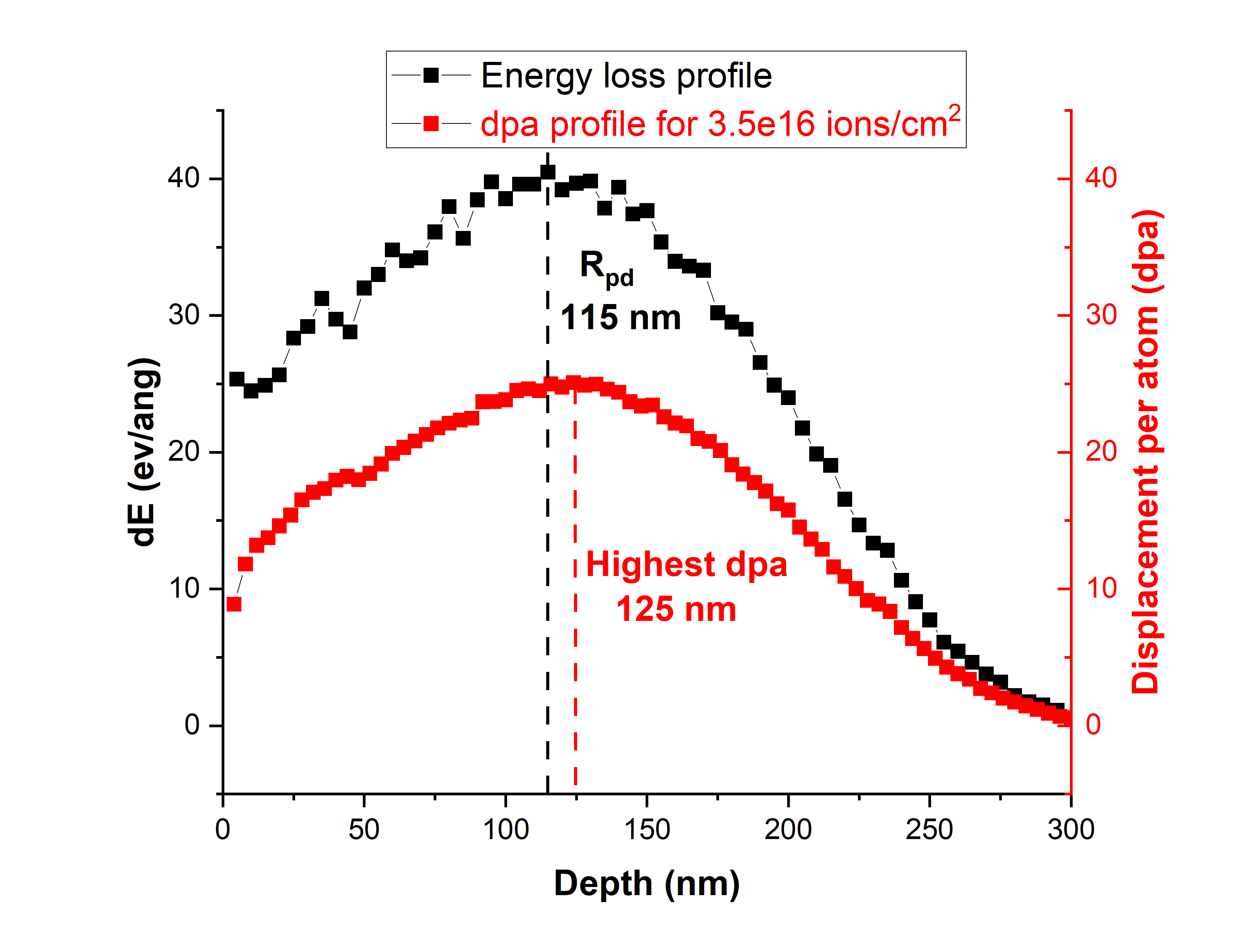}
    \caption{Calculated $R_\text{pd}$ and \ac{dpa} for \qty{140}{keV} Ne$^+$ irradiation of \BGaO{} using the SRIM code.}
    \label{fig:S10}
\end{figure}

$N_\text{vac}^\text{max}$  is the maximum value in the SRIM vacancy generation profile, while $n_\text{a}$ is the atomic density of \BGaO{}. 
Furthermore, we found that $R_\text{pd}$ (the depth of the maximum energy loss profile) is approximately \qty{115}{nm}, while the highest \ac{dpa} is at around \qty{125}{nm} as shown in Figure~\ref{fig:S10}. 
%-------------------------------------------------------------------
%-------------------------------------------------------------------

Therefore, as shown in the main text in Figure~\ref{fig:RBS}(b), an increase in the backscattering yield in the first \qty{150}{nm} is expected.

\subsection{Supplementary Note 3 (Positrons)}

\begin{figure}[btp]
    \includegraphics[width=1\linewidth]{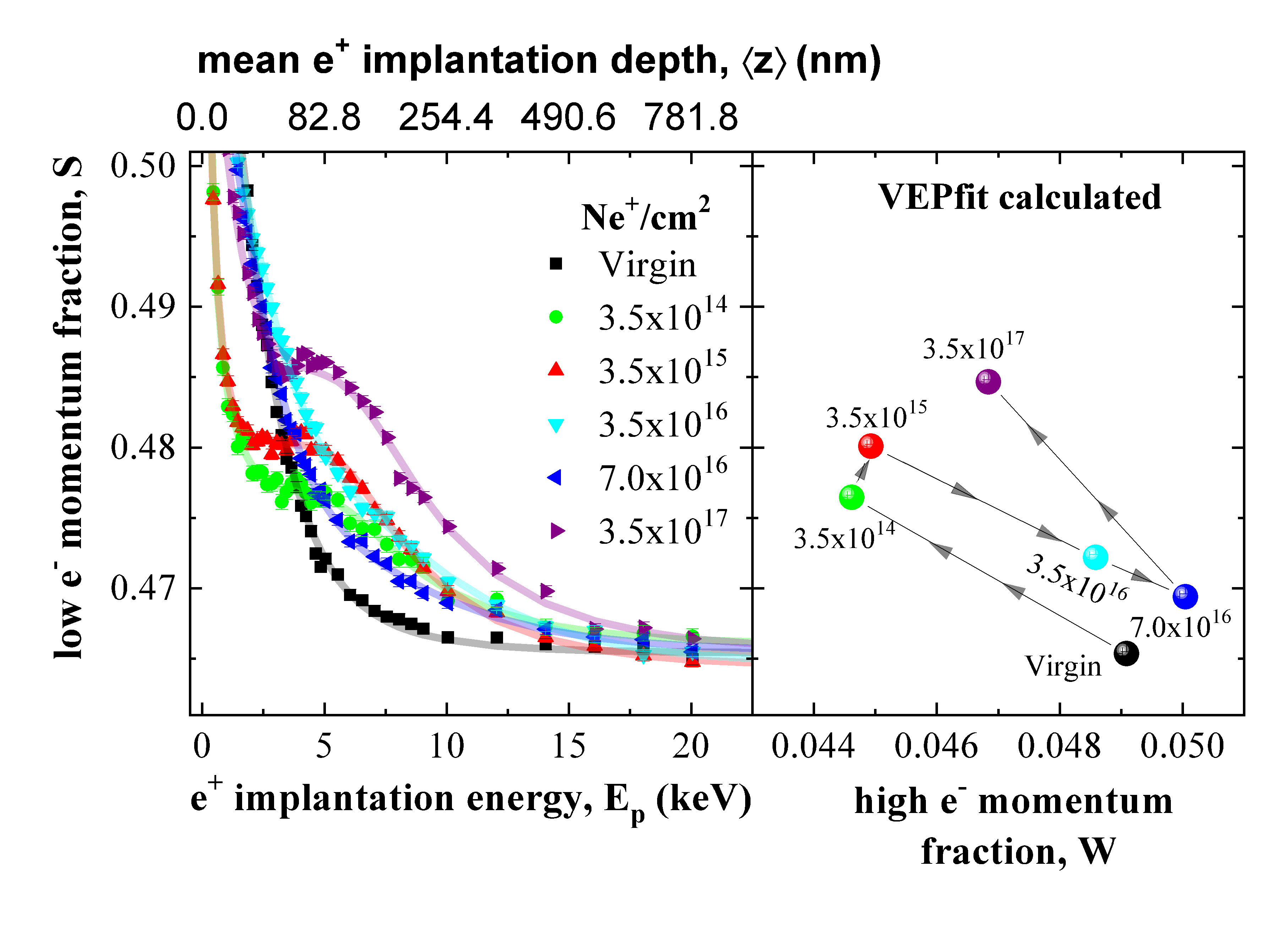}
    \caption{(left) S parameter vs positron implantation energy for each irradiation fluence on \BGaO{} and (right) corresponding \ac{vepfit}-calculated S-W fractions for each different irradiation fluences on \BGaO{} for \qty{5}{keV} positron implantation energy.}
    \label{fig:S11}
\end{figure}

In Figure~\ref{fig:S11}, the S-parameter is plotted as a function of the positron implantation energy $E_p$ for each irradiation fluence. 
%-------------------------------------------------------------------
%-------------------------------------------------------------------
The plateau observed in the S-parameter at certain energies corresponds to an increased defect concentration and defect accumulation within the material. 
\ac{vepfit} results are shown as calculated S-W fractions from the ion irradiated region. 
These calculations illustrate changes in the annihilation characteristics as the irradiation fluence increases. 
At lower fluences, such as \qty{3.5e14}{\ions\per\centi\meter\squared} and \qty{3.5e15}{\ions\per\centi\meter\squared}, the S-parameter increases, indicating a rise in the defect concentration due to the creation of new vacancy-type defects. 
This trend suggests that positrons are increasingly trapped at open-volume defects, such as monovacancies or small vacancy clusters. 
As the fluence increases to \qty{3.5e16}{\ions\per\centi\meter\squared}, a phase transition occurs in the material. 
This transition is marked by a reduction in the S-parameter and a corresponding increase in the W-parameter as well as the S($E_p$) curve decay without visible plateau, the analogue of the reference sample (Figure~\ref{fig:S11} left panel). 
The reduction in the S-parameter reflects a decrease in the concentration of open-volume defects and possibly a redistribution of defects due to the phase transition. 
At the highest fluence, \qty{3.5e17}{\ions\per\centi\meter\squared}, the material reaches its maximum S-parameter, indicating a significant accumulation of open-volume defects. 
This shows a highly damaged structure.

\begin{figure}[btp]
    \includegraphics[width=1\linewidth]{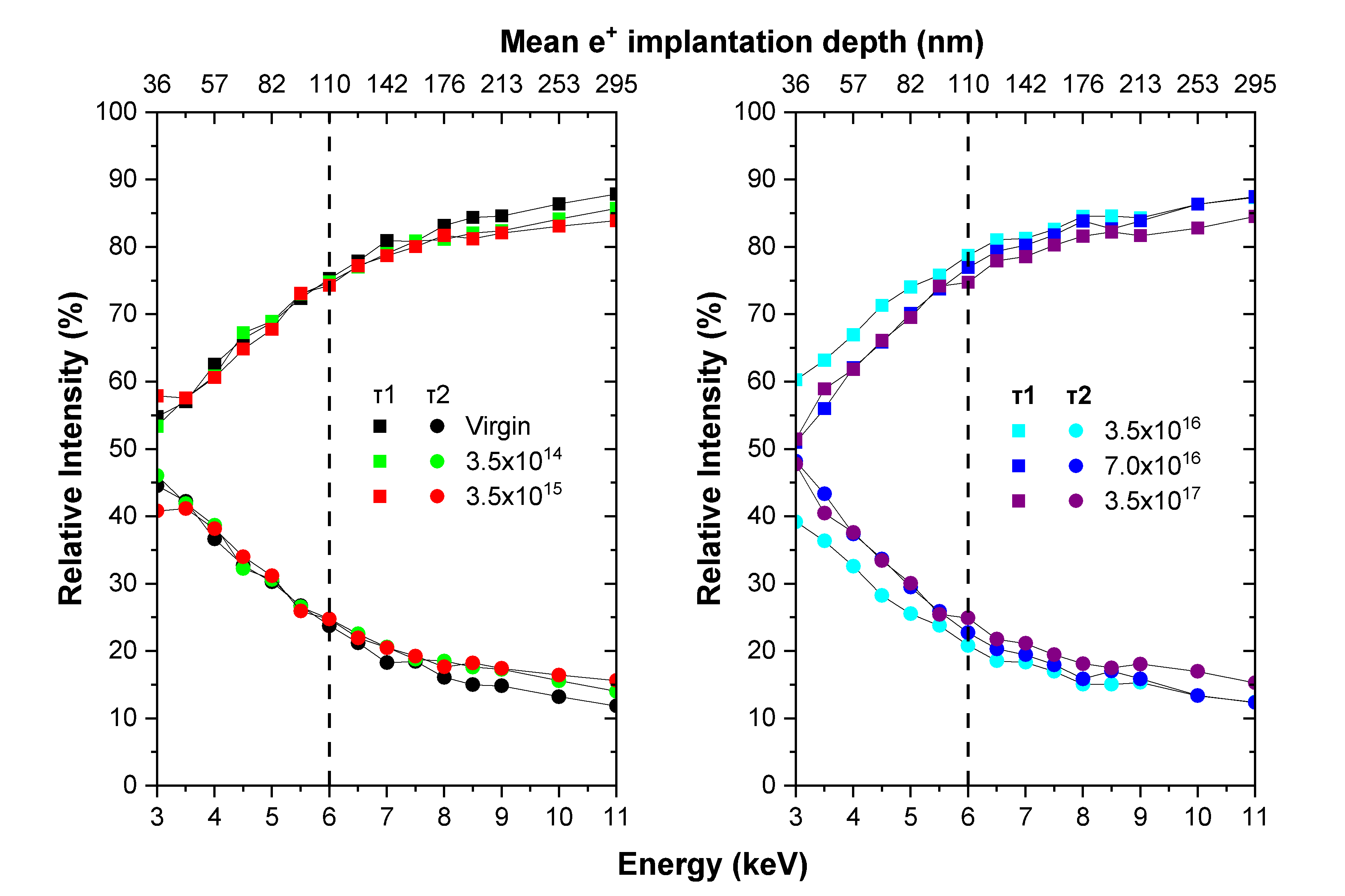}
    \caption{The relative intensities of the positron signals are plotted as a function of positron implantation energy for samples with different irradiation fluences. 
    The intensity values along the dashed lines, corresponding to a positron implantation energy of \qty{6}{keV}, were used in Figure~\ref{fig:ratiocurves} of the main text.}
    \label{fig:S12}
\end{figure}

In Figure~\ref{fig:S12}, the relative intensities for Ne$^+$-irradiated samples are plotted.
%-----------------------------------------------------------------
%-----------------------------------------------------------------
As observed, the positron signals are dominated by short-lived positrons ($\tau_1$). 
Therefore, in the main text, we focused on $\tau_1$ lifetimes. 
There are negligible changes in the relative intensities as the irradiation fluence increases, which is the consequence of positron saturation trapping.

\begin{figure}[btp]
    \includegraphics[width=1\linewidth]{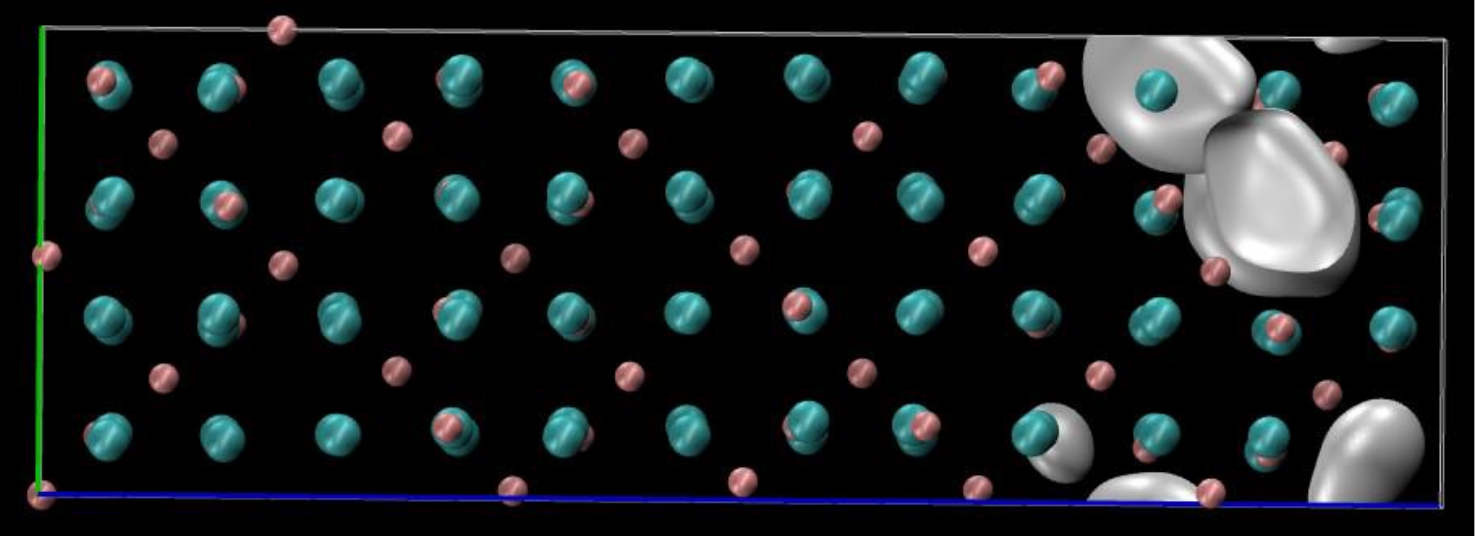}
    \caption{2-site \GGaO{} model (two maximum positron density points).}
    \label{fig:S13}
\end{figure}

\begin{figure}[btp]
    \includegraphics[width=1\linewidth]{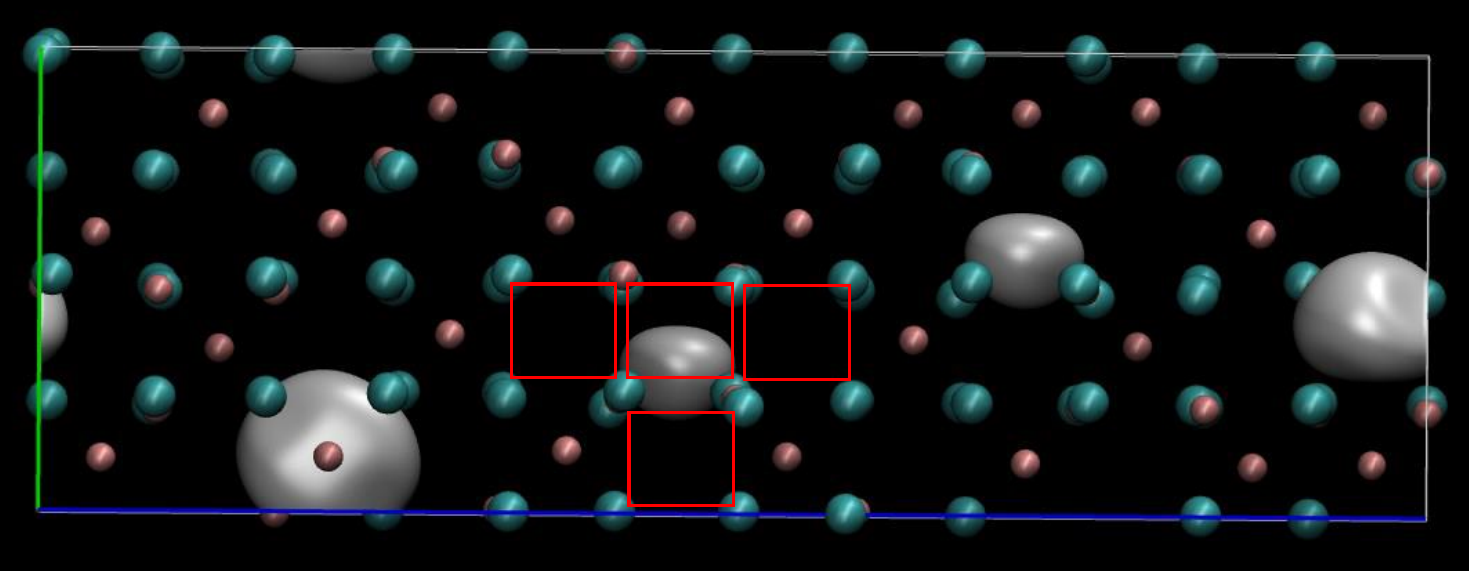}
    \caption{3-site \GGaO{} model with four maximum positron density points (four V$_{Ga}$ sharing one positron density max point, shown in red squares).}
    \label{fig:S14}
\end{figure}

\begin{figure}[btp]
    \includegraphics[width=1\linewidth]{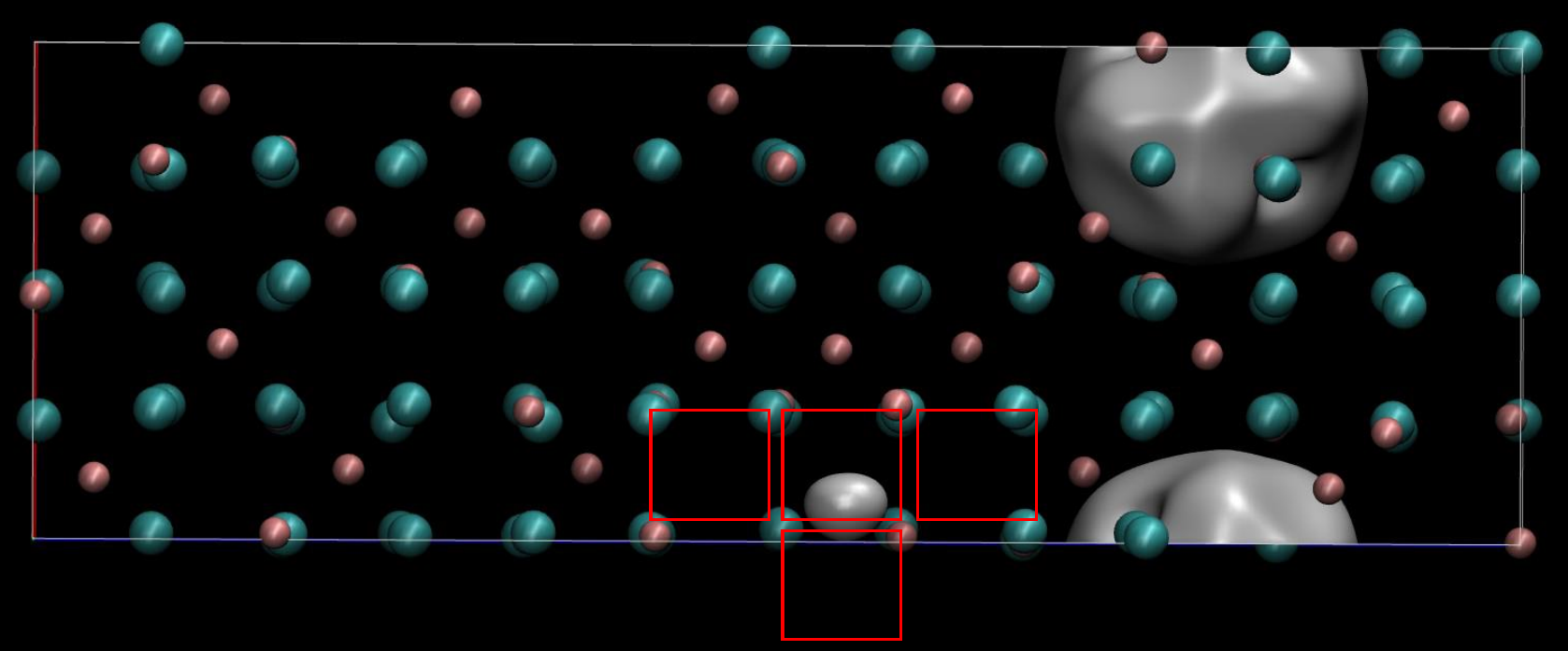}
    \caption{4-site \GGaO{} model with two maximum positron density points (four and five V$_{Ga}$ sharing one positron density max point, shown in red squares).}
    \label{fig:S15}
\end{figure}

In Figures~\ref{fig:S13}, \ref{fig:S14}, \ref{fig:S15} the maximum positron density points for different \GGaO{} models are presented.
%-----------------------------------------------------------------
%-----------------------------------------------------------------
%-----------------------------------------------------------------
%-----------------------------------------------------------------
%-----------------------------------------------------------------
%-------------------------------------------------------------------
Maximum positron density points refer to specific regions within the \GGaO{} structure where the density of positrons is highest. 
These points indicate areas where positrons are most likely to interact with electrons, leading to annihilation.

\begin{figure}[btp]
    \includegraphics[width=1\linewidth]{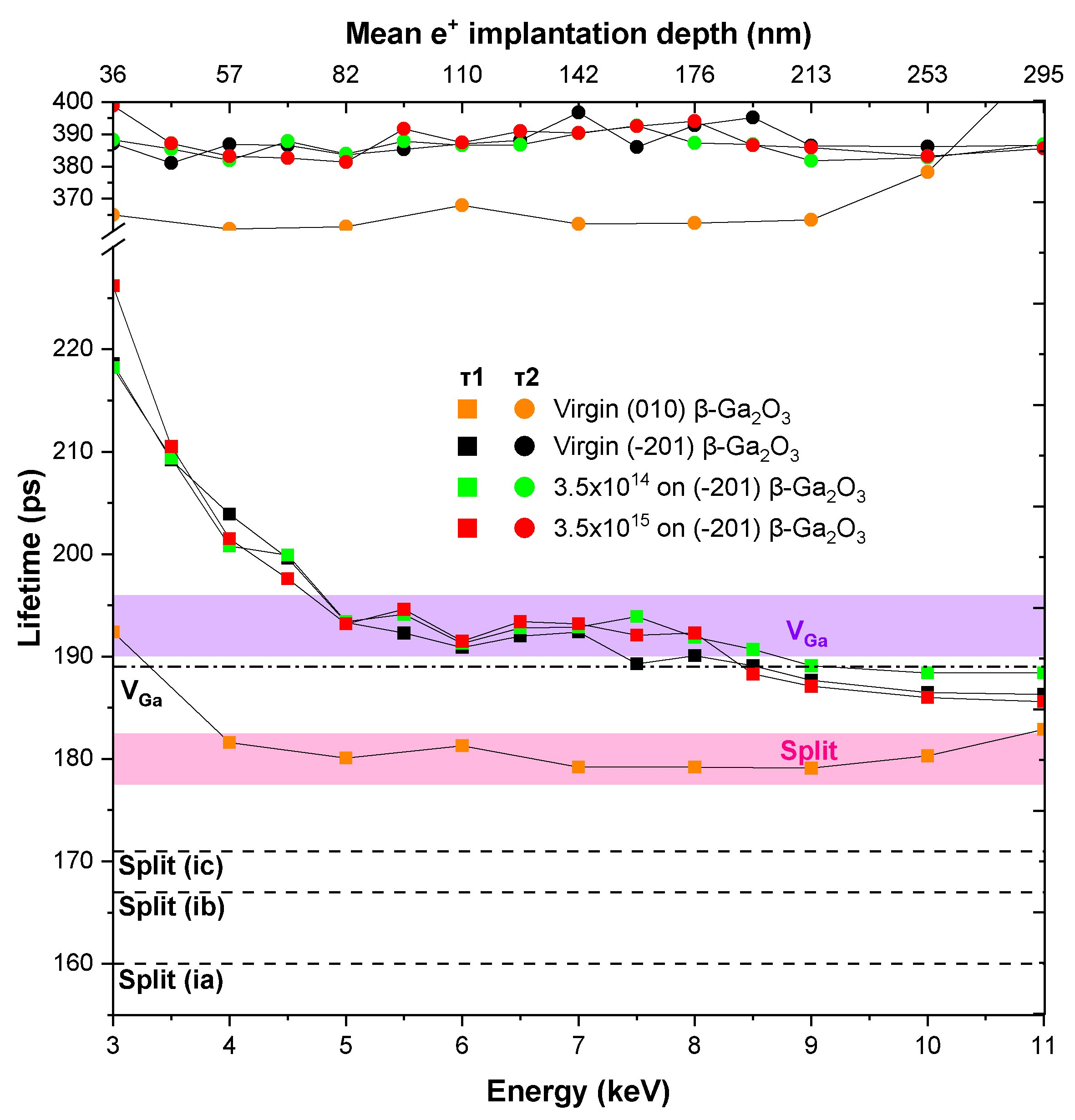}
    \caption{Positron lifetime of the \hkl(010)-oriented reference \BGaO{} sample compared to ($\overline{2}01$)-oriented reference \BGaO{} and \qty{140}{keV} Ne$^+$ irradiated samples.}
    \label{fig:S16}
\end{figure}

The size and number of maximum positron density points influence positron lifetime, as shown in Table~\ref{tab:S2}.
%----------------------------------------------------
\begin{table}[tbp]
  \caption{Calculated lifetime values for different \GGaO{} models. %\qty{135}{ps} bulk \BGaO{} value is used for comparison%~\cite{Karjalainen2020}.
For comparision the bulk \BGaO{} lifetime value of \qty{135}{ps} is given in the table~\cite{Karjalainen2020}.
  }
  \label{tab:S2}
  \centering
  \begin{tabular}{|l|l|l|l|}
    \hline
    \textbf{\GGaO{}}    & \textbf{2-Site} & \textbf{3-Site} & \textbf{4-Site} \\ \hline
    Lifetime (ps) & 164             & 172             & 173             \\ \hline
    $\delta$ (compare to \BGaO{})            & 29              & 37              & 38              \\ \hline
  \end{tabular}
\end{table}
%-----------------------------------------------
For comparision the bulk \BGaO{} lifetime value of \qty{135}{ps} is given in the table~\cite{Karjalainen2020}.
Higher lifetime values in \GGaO{} compared to \BGaO{} suggest that \GGaO{} contains larger size of positron traps.
We also measured the lifetime of the \hkl(010)-oriented \BGaO{} as shown in Figure~\ref{fig:S16} and observed split vacancy configuration in the subsurface region.
%-----------------------------------------------------------------
%----------------------------------------------------------------

The reason for observing two different defect configurations for \hkl(010) and ($\overline{2}01$)-oriented reference samples could be attributed to differences in the \ac{pad} between the ($\overline{2}01$) and \hkl(010) planes~\citeS{Yao2020}. 
Considering that slip planes in \BGaO{} are defined by crystallographic planes with high \ac{pad}~\citeS{Yao2020}, ($\overline{2}01$) and \hkl(010) planes exhibit different \ac{pad} values. 
Consequently, the strain and damage induced by surface preparation may affect the subsurface region differently for these two orientations, leading to variations in the defect density and configuration. 

\begin{figure}[btp]
    \includegraphics[width=1\linewidth]{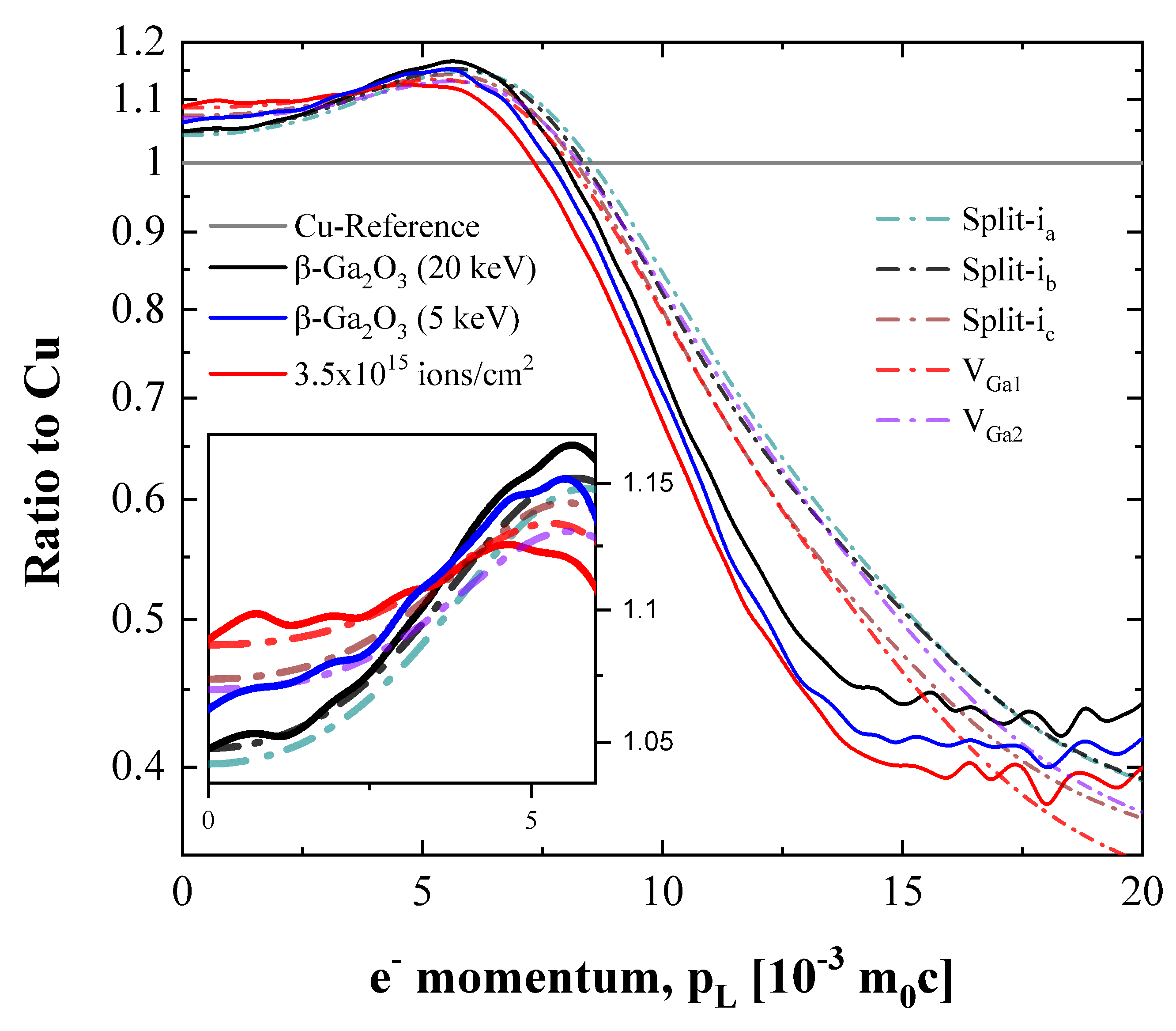}
    \caption{The ratio curves show the momentum distribution of annihilation electrons. 
    The dash-dotted lines represent the theoretically calculated momentum distribution for defect configurations in \BGaO{}.
    A positron implantation energy of \qty{20}{keV} and \qty{5}{keV} were used for the \BGaO{} reference samples, while \qty{5}{keV} was used for the irradiated sample.
    For clarity, the inset shows a magnified view of the low-momentum region.}
    \label{fig:S17}
\end{figure}

Furthermore, we applied \ac{cdb} spectroscopy to analyze the momentum distribution of electrons involved in annihilation. 
The \ac{cdb} ratio curves shown in Figure~\ref{fig:S17} are normalized to the momentum distribution of an annealed copper sample, which serves as a defect-free reference material with a high positron diffusion length.
%-----------------------------------------------------------------
%----------------------------------------------------------------
A positron implantation energy of \qty{20}{keV} was used for the reference sample to investigate annihilation events in deeper regions, while \qty{5}{keV} was used  to investigate annihilation events in the subsurface regions. 
According to our \ac{dft} calculations, the split vacancies exhibit lower (higher) intensities in the low (high) electron momentum regions compared to Ga-vacancies, especially of type one (V$_{Ga1}$). 
In the low-momentum region, the results for the reference sample measured at a positron implantation energy of \qty{20}{keV} indicate, on average, the split defect configuration rather than the V$_{Ga}$ defect configuration. 
This energy provides information about the annihilation events occurring deeper within the sample. 
Additionally, a \qty{5}{keV} implantation energy was used for the reference sample to investigate surface-related annihilation events. 
The results show an increased intensity in the low-electron-momentum region compared to the reference sample measured at \qty{20}{keV}, aligning with the V$_{Ga}$ (instead of the split defect configuration or larger vacancy clusters). 
The reduced intensity in the high-momentum region also indicates the presence of V$_{Ga}$ or larger vacancy clusters in the subsurface region. 
After irradiation with \qty{3.5e15}{\ions\per\centi\meter\squared}, the low-momentum region exhibits an increase, aligning with the V$_{Ga}$. 
The reduced intensity in the high-momentum region also aligns with the trend calculated via \ac{dft} (dashed lines) and suggests the presence of V$_{Ga}$. 
The data presented here are in excellent agreement with Figure~\ref{fig:PALS}(b) of the main text.

\clearpage
%\bibliographystyleS{MSP}
\bibliographystyleS{plain}
\bibliographyS{Defects-in-Ga2O3_v2}

\end{document}